\documentclass[a4paper, 11pt]{article}
\usepackage[english]{babel}
\usepackage{setspace}
\usepackage{graphicx}
\usepackage{float}
\usepackage{enumerate}
\usepackage{url}
\usepackage{amsmath}
\usepackage[super,numbers,sort&compress]{natbib}
\usepackage{hyperref}
\usepackage{cleveref}
\usepackage[T1]{fontenc}
\usepackage{verbatim}
\usepackage{braket}
\usepackage[affil-it]{authblk}
\usepackage[textheight=23cm,textwidth=17cm]{geometry}

\author{Lennard B\"oselt}
\author{Moritz Th\"urlemann}
\author{Sereina Riniker\thanks{Corresponding author: email: sriniker@ethz.ch, ORCID: 0000-0003-1893-4031}}
\affil{Laboratory of Physical Chemistry, ETH Zurich, Vladimir-Prelog-Weg 2, 8093 Zurich, Switzerland}

\title{\textbf{Machine Learning in QM/MM Molecular Dynamics Simulations of Condensed-Phase Systems}}

\date{}

\begin{document}

\maketitle
\begin{abstract}
Quantum mechanics/molecular mechanics (QM/MM) molecular dynamics (MD) simulations have been developed to simulate molecular systems, where an explicit description of changes in the electronic structure is necessary. However, QM/MM MD simulations are computationally expensive compared to fully classical simulations as all valence electrons are treated explicitly and a self-consistent field (SCF) procedure is required. Recently, approaches have been proposed to replace the QM description with machine learned (ML) models. However, condensed-phase systems pose a challenge for these approaches due to long-range interactions. Here, we establish a workflow, which incorporates the MM environment as an element type in a high-dimensional neural network potential (HDNNP). The fitted HDNNP describes the potential-energy surface of the QM particles with an electrostatic embedding scheme. Thus, the MM particles feel a force from the polarized QM particles. To achieve chemical accuracy, we find that even simple systems require models with a strong gradient regularization, a large number of data points, and a substantial number of parameters.
To address this issue, we extend our approach to a $\Delta$-learning scheme, where the ML model learns the difference between a reference method (DFT) and a cheaper semi-empirical method (DFTB). We show that such a scheme reaches the accuracy of the DFT reference method, while requiring significantly less parameters. Furthermore, the $\Delta$-learning scheme is capable of correctly incorporating long-range interactions within a cutoff of 1.4\,nm. It is validated by performing MD simulations of retinoic acid in water and the interaction between S-adenoslymethioniat with cytosine in water. The presented results indicate that $\Delta$-learning is a promising approach for (QM)ML/MM MD simulations of condensed-phase systems. 
\end{abstract}

\section{Introduction}
Classical fixed-charge force fields (FF) are readily used to perform molecular dynamics (MD) simulations of condensed-phase systems \cite{Riniker2018,Nerenberg2019}. They consist of a relatively small number of parameters, such as bond-stretching, bond-angle bending, and torsional dihedral terms, partial charges and Lennard-Jones parameters. They are partially fitted against experimental values such as the density, heat of vaporisation, and solvation free energy.\cite{Schmid2011,Ponder2013}
FF are the gold standard for simulations over long time scales of systems for which long-range interactions are essential. In classical simulations, averaged properties are computed by neglecting electron rearrangement. The parameterization of FF requires the availability of sufficient experimental data.
On the other hand, quantum mechanics/molecular mechanics (QM/MM)\cite{Warshel1976,Mulholland2000,Thiel2009,Groenhof2013} MD simulations can provide a valuable alternative to classical FF simulations, when changes in the electronic structure are important or if reliable force-field parameters are not available.

In the QM/MM scheme, the QM zone simulated with density functional theory (DFT) or \textit{ab initio} principles is placed into a classical environment (MM zone). This approach permits the simulation of the electronic structure of small systems in more realistic surroundings. Most crucial in QM/MM simulations is the description of the interaction between the QM and MM zones. This interaction can be based either on (i) mechanical constraints (i.e. ``mechanical embedding" scheme), or (ii) electronic perturbations (i.e. ``electrostatic embedding''). In mechanical embedding, the particles in the QM zone are assigned partial charges, which then interact with the MM zone on a classical level. Thus, the MM particles do not interact with the electron density of the QM solute but rather with point charges on a classical level. Mechanical embedding favours efficient implementations, however, a major drawback is that a set of classical parameters have to be determined for the QM zone, which is not trivial and sometimes not possible with sufficient accuracy.\cite{Lin2007}
In electrostatic embedding, the MM environment is incorporated into the Hamiltonian operator of the QM zone as electron operators. Thus, the electron density of the QM zone is perturbed by the MM environment, and the MM zone in turn interacts with the perturbed electronic structure of the QM solute. In other words, the MM particles feel a force from the MM-polarized QM solute. 
The advantages of electrostatic embedding are that no partial charges have to be assigned for the computation of the interactions between the MM and the QM zones, and that the description of the interaction is physically better motivated compared to mechanical embedding. A well known limitation is the neglect of polarization of the MM environment -- unless polarizable FF are used\cite{Loco2017}.
Furthermore, computation is more expensive compared to mechanical embedding, and it is not clear whether the partial charges of the MM zone are suitable for inclusion as one-electron Hamiltionans in the QM zone.\cite{Lin2007,Panosetti2019} Generally, an electrostatic embedding scheme is to be preferred as it has been shown to be more accurate, and it reflects thus the current standard protocol for QM/MM simulations.\cite{Lin2007,Thiel2009,Brunk2015,Chung2015,Panosetti2019}

In QM/MM, the description of the QM zone becomes the computational bottle-neck as it requires an expensive self-consistent field (SCF) procedure, and explicit treatment of all valence electrons\cite{Neese2018,Balasubramani2020}. To partially circumvent these issues, semi empirical methods\cite{Brunk2015,Christensen2016} can be used to describe the QM zone\cite{Stewart2013}. This extends the accessible time scales but also reduces the accuracy.\cite{Brunk2015} An alternative is to employ machine-learned (ML) potentials.

In recent years, a lot of research effort has been invested into the development of ML models trained on the potential energy surface (PES) of QM systems\cite{Behler2011,Bin2013,Li2013,Behler2014,Behler2015,Botu2015,Ramakrishnan2015,Gastegger2017,Gastegger2018,chmiela2017,Chmiela2018,Smith2017,Noe2020}. This enables ML-MD simulations at an accuracy level close to that of the electronic structure method chosen to generate the training set. The costs of the resulting ML-MD simulations are reduced drastically compared to a normal DFT or \textit{ab initio} MD simulation as a SCF procedure is no longer necessary and the valence electrons do not have to be treated explicitly. A large amount of approaches have been reported in the literature to achieve this task. These approaches differ in: 
\begin{itemize}
  \item \textit{Use case}: e.g. gas-phase\cite{Bin2013,Li2013} or periodic box\cite{Behler2011}, small compounds\cite{Li2013} or larger compounds,\cite{Smith2017} system specific\cite{chmiela2017,Chmiela2018} or system unspecific\cite{Smith2017}
  \item \textit{Descriptor used to encode the chemical structure}: e.g. Coulomb matrices\cite{Pereira2012}, distance matrices\cite{chmiela2017,Chmiela2018}, low-order polynomials,\cite{Bin2013,Li2013} or so-called symmetry functions\cite{Behler2011,Botu2015,Smith2017}
  \item \textit{Target output}: e.g. energies\cite{Behler2011,Bin2013} or gradients\cite{chmiela2017,Chmiela2018,Botu2015}
  \item \textit{Scope}: e.g. global\cite{chmiela2017,Chmiela2018} or local\cite{Behler2011,Behler2014,Behler2015,Smith2017,Botu2015}
  \item \textit{ML method}: e.g. neural networks\cite{Behler2011,Bin2013} or Kernel methods\cite{chmiela2017,Chmiela2018,Brunken2020}
  \item \textit{Applying a restriction of the final form of the PES}\cite{Brunken2020}
  \item \textit{Convolution of the chemical graph}: Using a set of hard-coded functions,\cite{Behler2011,chmiela2017,Bin2013} or using other techniques such as graph (convolutional) architectures\cite{SchNet}
\end{itemize}
Other ML models do not target the PES but rather the electron density\cite{Fabrizio2019} or the wave function itself.\cite{OrbNet} However, these will not be subject of the current study. All of the approaches listed above partially generate the correct parity behaviour of the system. In other words, swapping two elements of the same type and rotating or translating the system leaves the output unchanged or changes it equivariantly. The usage of the ML models heavily depends on the specific use case. In the following, selected approaches are discussed in more detail.

Permutational-invariant polynomial neural networks (PIP)\cite{Bin2013} are suitable for accurately describing reactions of small systems in the gas phase (less than 5 atoms). Here, low-order polynomials, which depend on the internuclear distance of the compounds, are used as a descriptor for rigorously introducing the correct parity behaviour for the system studied. The neural networks are trained on the total energy of the system and are global. 
Symmetric gradient domain machine learning (sGDML)\cite{chmiela2017,Chmiela2018} targets systems up to 30 atoms in the gas phase, which do not undergo a change in topology. In this case, the compound is encoded as an inverse distance matrix. The latter is fed into a Kernel ridge regression model and trained in the gradient domain, i.e. the target output is the gradient of the system. The energy can be obtained by integrating the forces. The definition of the kernel used ensures that the integrated forces give rise to a continuously-differentiable PES. The models can be employed for spectroscopic studies\cite{Chmiela2018}. 

Larger compounds and box simulations can be tackled with high-dimensional neural network potentials (HDNNP)\cite{Behler2011,Behler2014,Behler2015}. The environment of each atom is encoded using symmetry functions. Simple examples of symmetry functions are Gaussians that depend on the interatomic distance, or trigonometric functions. They map the input to a higher dimension, which is subsequently summed up. The summation operator and the mapping to higher dimensions using symmetry functions introduces the correct parity behaviour,\cite{sannai2020} while still allowing the HDNNP to learn the PES accurately. Using symmetry functions, Smith \textit{et al.}\cite{Smith2017} showed that a universal force-field (ANI-1) can be constructed, which is capable of describing unseen compounds containing the same elements. The model was trained in a system unspecific manner on gas-phase data. The models can be employed in the drug-discovery process.\cite{Stevenson2019} 
For periodic-box simulations, HDNNP can be trained system-specific on previously sampled data points.\cite{Behler2011,Behler2014,Behler2015} The sampling of the data points can be done on a lower level of theory, as long as the trajectory samples all structures important for learning the PES. The data points are then re-computed on the desired level, and the HDNNP is trained to interpolate between the training data points. These models employ a cutoff for the symmetry functions and are thus local. They are trained on the energy and use optionally a gradient correction.\cite{Behler2015} 
HDNNP based on symmetry functions motivated other groups to apply a similar approach for different target outputs such as gradients, or with different ML models such as Kernel methods.\cite{Botu2015}. The uniqueness of the descriptor has been discussed in theoretical work.\cite{Pozdnyakov2020} 

Another important contribution to the field is the concept of $\Delta$-learning.\cite{Ramakrishnan2015} Here, the ML model is trained to reproduce the difference between two QM calculations, an expensive, higher-level and a cheaper, lower-level method. Different ML methods and descriptors can be used. Subsequently, the output on the higher level can be partially recovered by performing the calculation with the cheap method and applying the trained ML model. A good example of the $\Delta$-learning scheme was proposed by Shen \textit{et al.},\cite{Shen2016} where a modified HDNNP was trained on the energy difference between a semi-empirical and an expensive QM method. It was used to reweight the free-energy profiles of reactions obtained with a QM/MM approach. More recently, the same authors introduced a model, which can be used to perform MD simulations. The approach requires as input for the ML model a reaction coordinate and the partial charges from the lower level method.\cite{Shen2018}

In our opinion and that of others,\cite{Noe2020} HDNNP have been proven to be currently the most suitable method for performing system-specific periodic-box simulations. However, their usage in the simulation of condensed-phase systems with biological relevance is still hampered because such systems typically involve:\cite{Zhang2018,Panosetti2019,Gastegger2018,Noe2020,Brunken2020}
\begin{enumerate}
\item \textit{Large number of element types}: HDNNP input descriptors scale exponentially with the number of element types.
\item \textit{Many rotatable bonds and/or no symmetry, large system size}: The quality of HDNNP model depends on a densely sampled training set. The number of different possible system configuration scales exponentially with the number of rotatable bonds, which renders sufficient sampling for the training set increasingly challenging. In principle, if the ML model is able to learn the underlying physics, it could extrapolate to unseen system configurations. However, to the best of our knowledge, this was only partially achieved yet. The models, which achieve this task partially, employ a small cutoff for the many-body (0.32\,nm) and two-body terms (0.52\,nm).\cite{Smith2017,Stevenson2019}
\item \textit{Important long-range interactions}: HDNNP make use of a locality ansatz\cite{Grisafi2019}, which fails to describe long-range electrostatics. Long-range interactions can partially be introduced via a charge partition scheme and a charge summation scheme (e.g. Ewald summation).\cite{Behler2015} In QM/MM, this would correspond to a mechanical embedding scheme.\cite{Junming2018} ML approaches have been proposed to specifically target this issue,\cite{Grisafi2019,Pozdnyakov2020} but these are to the best of our knowledge not yet applicable to simulations of large condensed-phase systems.
\item \textit{Large cutoff}: Classical force fields typically work with a cutoff of 1.0 - 1.4 nm for pairwise nonbonded interactions.\cite{Schmid2011,Riniker2018,Nerenberg2019} Such large cutoffs are necessary to achieve the desired accuracy for bulk properties such as heat of vaporization or solvation free energy.\cite{Junming2018}
\item \textit{Long time scales}: Although a prediction with the ML model is faster than a QM calculation, a small integration step (i.e. 0.5\,fs) is still required for simulations. Furthermore, obtaining the gradients from a ML potential is usually significantly slower than calculating the gradients with a classical force field.\cite{Stevenson2019}
\end{enumerate}
In addition, HDNNP and other ML models are generally plagued by two issues: (i) Generating training sets can imply a significant time investment, and (ii) frequently used fully connected neural networks are prone to overfitting\cite{Behler2017}, especially if a high number of weights is necessary to converge to chemical accuracy.\cite{MorphNet}

A promising alternative for condensed-phase systems is therefore the combination of HDNNP and classical FF in a QM/MM-type approach, where the HDNNP is used for simulating the QM particles and to compute the interactions between the MM and the QM zone with electrostatic embedding. By using HDNNP for the QM zone, longer time scales will become accessible than with standard QM/MM simulations. Such a hybrid approach requires the training data set to be generated with a QM/MM scheme. In other words, the MM environment is incorporated as one-electron Hamiltonians in the QM reference calculation. With such an approach, the complexity of the generation of the training set is reduced drastically as not all atoms are treated on a QM level. Furthermore, it permits the use of a larger cutoff. Including long-range interactions directly in the ML model is therefore possible.

In this work, we assess the concept of (QM)ML/MM MD simulations for condensed-phase systems. To enable electrostatic embedding, we integrate the MM environment directly as an additional element type in the HDNNP. The symmetry functions including MM particles are weighted with their respective partial charge. Two-body and three-body symmetry functions are included for the description of the MM environment. The symmetry functions used follow the proposal of Smith \textit{et al.}\cite{Smith2017,Stevenson2019} and are adapted for larger cutoffs. 
In the first part, we explore the effect of different parameters on the accuracy of (QM)ML/MM calculations of small molecules in water. The parameters include the application of gradient correction, the inclusion of three-body terms, and the cutoff distance for the descriptors.
In the second part, we extend the methodology to a $\Delta$-learning scheme\cite{Ramakrishnan2015} with a semi-empirical method. Semi-empirical methods are known to recover long-range interactions well\cite{Panosetti2019}, an issue ML models with a locality ansatz usually struggle with. The approach is validated using different molecular systems. Finally, MD simulations are performed for retinoic acid in water (50 atoms in the QM zone and approximately 2500 classical partial charges), and the interaction of SAM with cytosin (63 atoms in the QM zone and approximately 3000 partial charges). The results are compared to standard QM/MM simulations.

\section{Theory}
\subsection{Calculation of the Potential Energy of the System}
\subsubsection{Choice of the Model}
In non-relativistic Born-Oppenheimer approximated,\cite{Born} time-independent wave function mechanics\cite{Reiher2014}, the electronic energy including nuclear repulsion $E_\text{QM}$ can be computed by solving the Eigen function,\cite{Schroed1926}
\begin{equation}
    \hat{H}_\text{QM}\psi_{\vec{R}} (\vec{r}) = E_{\text{QM}}(\vec{R})\psi_{\vec{R}}(\vec{r}).
    \label{eq:SG}
\end{equation}
In Eq.~(\ref{eq:SG}), $\psi_{\vec{R}}(\vec{r})$ is the electronic wave function, $\vec{r}$ are the coordinates of the electrons, and $\vec{R}$ the coordinates of the nuclei. The subscript $\vec{R}$ denotes that the wave function depends on the coordinates of the nuclei as a parameter. We will use the notation $\psi_{\vec{R}}(\vec{r})\equiv \psi(\vec{r})$ in the following. $\hat{H}_{\text{QM}}$ is the electronic Hamiltonian operator including the nuclear core-core repulsion and reads in atomic units,
\begin{equation}
    \hat{H}_\text{QM}=-\frac{1}{2}\sum_i^{N_\text{el}} \nabla_i^2 + \sum_{i<j}^{N_\text{el}} \frac{1}{|\vec{r}_i - \vec{r}_j|} - \sum_{i}^{N_\text{el}}\sum_{j}^{N_\text{QM}} \frac{Z_{j}}{|\vec{r}_i-\vec{R}_j|}+\sum_{i<j}^{N_\text{QM}} \frac{Z_{i}Z_{j}}{|\vec{R}_{i}-\vec{R}_j|} ,
    \label{eq:expectation}
\end{equation}
where $Z_j$ is the charge of the nuclei $j$. The indices run over the total number of electrons $N_\text{el}$ and the total number of nuclei $N_\text{QM}$.
The quantum Born-Oppenheimer approximated potential energy $E_{\text{QM}}$ is obtained by computing the expectation value,
\begin{equation}
    E_{\text{QM}}(\vec{R}) = \frac{\braket{\psi(\vec{r})|\hat{H}_\text{QM}\psi(\vec{r})}}{\braket{\psi(\vec{r})|\psi(\vec{r})}}.
\end{equation}
According to Newton's second law,\cite{Newton} the gradients on the nuclei of the system can be computed by taking the derivative with respect to the coordinates $\vec{R}$,\cite{Reiher2014}
\begin{equation}
    -\frac{\partial E_{\text{QM}}(\vec{R})}{\partial \vec{R}_{i}} = \vec{F}_{i} = m_{i}\frac{\text{d}\vec{v}_{i}}{\text{d}t}.
\label{eq:Newtonslaw}
\end{equation}
In Eq.~(\ref{eq:Newtonslaw}), $\vec{F}_{i}$ is the force of particle $i$, $m_{i}$ its mass, $\vec{v}_{i}$ its velocity, and $t$ is the time. The system can then be propagated in time $t$ using an integration algorithm such as the leap-frog algorithm.\cite{Gray1994} Propagating the system using the quantum Born-Oppenheimer approximated potential-energy surface is computationally unfeasible for large systems. Hence, approximated solutions to Eqs.~(\ref{eq:SG}) and (\ref{eq:expectation}) were developed, which can be grouped into four classes:\cite{Levine2013}
(1) \textit{Ab initio} methods use the correct Hamiltonian and an approximated wave function $\psi(\vec{r})$.  
(2) Density functional theory\cite{Kohn} (DFT) computes the electron density $\rho(\vec{r})$, which is used to calculate the energy of the molecular system, i.e. $E[\rho({\vec{r}})]$. Modern DFT approaches achieve an accuracy comparable to $ab$ $initio$ methods (or above), while still being able to simulate larger systems. However, DFT requires a treatment of all valence electrons and an SCF procedure, which makes it unfeasible to perform long MD simulations of large systems. 
(3) Semi-empirical methods use an approximated Hamiltonian $\hat{H}$ and correct for the error made by introducing a set of fitted parameters. Most prominent example is the PM7 method\cite{Stewart2013}. Another attractive alternative is density functional tight binding (DFTB),\cite{DFTB} which can be considered a semi-empirical variant of DFT. In DFTB, the electron density $\rho$ is expanded in a series around a reference density $\rho_{0}$ and truncated. The error made is corrected with fitted parameters, similarly to other semi-empirical methods.
(4) The last class includes empirical models with effective parameters such as classical force fields.\cite{Riniker2018,Christensen2016,Nerenberg2019}

\subsubsection{Classical Force Fields}
In classical fixed-charge MD simulations, the potential energy of the system is calculated with a force field,\cite{Riniker2018,Christensen2016,Nerenberg2019} which is the sum of bonded and nonbonded interactions terms,
\begin{equation}
    E_{\text{MM}}(\vec{R}) = E^{\text{bond}}(\vec{R})+E^{\text{angle}}(\vec{R})+E^{\text{dihedral}}(\vec{R})+E^{\text{el}}(\vec{R})+E^{\text{vdW}}(\vec{R}),
\label{eq:FF}
\end{equation}
where $E^{\text{bond}}$ is the contribution of all covalent bonds, $E^{\text{angle}}$ that of all covalent angles, and $E^{\text{dihedral}}$ that of all covalent dihedrals. The nonbonded terms consist of electrostatic ($E^{\text{el}}$) and van der Waals ($E^{\text{vdW}}$) interactions. In many force fields, the van der Waals interactions are only calculated up to a cutoff and neglected afterwards (straight truncation).\cite{Riniker2018} To indicate that only short-range van der Waals interactions are included, we will use the notation $E^{\text{vdW,SR}}$ in the following. The parameters of these interactions terms are typically fitted to reproduce experimental and/or QM reference data. Eq. (\ref{eq:FF}) is inexpensive to solve, and depends only on the number of nuclei (atoms). The inclusion of experimental data in the fitting procedure allows to incorporate long range effects in an implicit manner. Thus, they are capable of reproducing bulk properties, such as the solvation free energy accurately (see e.g. Refs.~\citenum{Mobley2014,Augustinus2017,Kashefolgheta2020}). However, fixed-charge force fields treat electronic effects only in an averaged field, and can thus not be used to simulate chemical reactions. The simplicity of Eq.~(\ref{eq:FF}) allows to use relatively large cutoffs, and methods are available to include long-range interactions beyond a given cutoff.

\subsubsection{QM/MM Scheme}
QM/MM is a hybrid approach that combines a QM subsystem with an MM environment,\cite{Warshel1976,Mulholland2000,Thiel2009,Groenhof2013} thus striking a balance between cost and accuracy. The challenge is to describe the interactions between the parts appropriately.
Here, the question is how to obtain the energy $E_{\text{QM/MM}}$ of the system, which is now a combination of the QM subsystem $E_{\text{QM}}$ and the MM surrounding $E_{\text{MM}}$. A first approach to compute the total energy of the system is a subtractive scheme,\cite{Thiel2009}
\begin{equation}
    E_{\text{QM/MM}}(\vec{R}) = E_{\text{QM}}(\vec{R}_\text{QM}) + E_{\text{MM}}(\vec{R}) - E_{\text{MM}}(\vec{R}_{\text{QM}}),
\label{eq:sub}
\end{equation}
where $E_{\text{QM}}(\vec{R})$ is the energy of the QM subsystem, $E_{\text{MM}}(\vec{R})$ the energy of the complete system calculated with the classical force field, and $E_{\text{MM}}(\vec{R}_{\text{QM}})$ the energy of the QM zone calculated with the force field. Note the distinction between $\vec{R}$, which refers to all nuclei in the system, and $\vec{R}_{\text{QM}}$, and $\vec{R}_{\text{MM}}$, which refer to the nuclei treated quantum-mechanically (QM zone) and the atoms treated classically (MM zone), respectively. 
The more prominent alternative is the additive scheme,\cite{Thiel2009} which is also used in this work, i.e.,
\begin{equation}
    E_{\text{QM/MM}}(\vec{R})= E_{\text{QM}}(\vec{R}_{\text{QM}}) + E^{\text{el}}_{\text{QM--MM}}(\vec{R})+ E^{\text{vdW,SR}}_{\text{QM--MM}}(\vec{R}) + E_{\text{MM}}(\vec{R}_{\text{MM}}).
\label{eq:add}
\end{equation}
Here, the superscript ``el'' stands for the electrostatic interactions, and the superscript ``vdW,SR'' for the short-range van der Waals interactions.
In this case, the force field is used to calculate the energy of the MM region, while the interaction between the QM and MM parts, $E^\text{el}_{\text{QM--MM}}$ and $E^\text{vdW,SR}_{\text{QM--MM}}$, is added to the energy expression. The latter term can be computed either via a mechanical embedding or an electrostatic embedding scheme.\cite{Lin2007} In mechanical embedding, the particles in the QM subsystem are assigned partial charges, which then interact with the MM surrounding. However, as already discussed in the Introduction, this scheme is inferior to electrostatic embedding. In electrostatic embedding, two Hamiltonians are introduced for dealing with the interactions between the MM environment and the QM particles. In atomic units, $\hat{H}_{\text{QM--MM}}^{\text{el}}$ reads,\cite{Thiel2009}
\begin{equation}
    \hat{H}^\text{el}_{\text{QM--MM}} = -\sum_{i}^{N_\text{MM}}\sum^{N_\text{el}}_{j}\frac{q_{i}}{|\vec{R}_{\text{MM},i}-\vec{r}_j|}+\sum_{i}^{N_\text{QM}}\sum_{j}^{N_\text{MM}}\frac{Z_{i}q_{j}}{|\vec{R}_{\text{QM},i}-\vec{R}_{\text{MM},j}|},
\label{eq:QMMMH}
\end{equation}
where $N_\text{MM}$ is the number of partial charges, and $q_j$ the partial charge of MM atom $j$. The QM subsystem is thus directly influenced by the MM partial charges, while the latter particles feel a force of the perturbed QM zone.  $\hat{H}^{\text{SR}}_\text{QM/MM}$ deals with the short-range van der Waals interaction, and is treated classically. It reads as follows,
\begin{equation}
    \hat{H}^{\text{vdW,SR}}_\text{QM--MM}= E^\text{vdW,SR}_\text{QM--MM}(\vec{R}) = \sum_{i}^{N_{\text{QM}}}\sum_{j}^{N_{\text{MM}}} 4 \epsilon_{ij} \Big(\Big(\frac{\sigma_{ij}}{|\vec{R}_i-\vec{R}_j|}\Big)^{12}-\Big(\frac{\sigma_{ij}}{|\vec{R}_i-\vec{R}_j|}\Big)^6\Big),
    \label{eq:SR}
\end{equation}
where $\epsilon_{ij}$ and $\sigma_{ij}$ are fitted parameters. 
Thus, in electrostatic embedding Eq.~(\ref{eq:add}) becomes,
\begin{equation}
    E_{\text{QM/MM}}(\vec{R}) =
    \frac{\braket{\psi(\vec{r})|(\hat{H}_{\text{QM}}+\hat{H}^{\text{el}}_{\text{QM--MM}})\psi(\vec{r})}}{\braket{\psi(\vec{r})\psi(\vec{r})}} +E_\text{QM--MM}^{\text{vdW,SR}}(\vec{R})
    +E_{\text{MM}}(\vec{R}_{\text{MM}})     .
\end{equation}

Typically, not all MM partial charges are included in the summation in Eq. (\ref{eq:QMMMH}), but only those within a cutoff radius $R_c$.\cite{Thiel2009} This leads to a PES, which is non-continuous at the cutoff. This issue can be (partially) resolved by adaptive resolution schemes.\cite{Bulo2009} It has been found that the cutoff radius $R_c$ needs to be chosen relatively large (i.e. 1.4\,nm) to converge to accurate results.\cite{Panosetti2019}

\subsection{Machine Learning of Potential-Energy Surfaces}
Supervised ML methods can be used to learn a function $f_P$\cite{Alpaydin2014,Paliwal2009,Kanagawa2018}, which describes the relation
$f_P:h(\vec{R})\rightarrow \tilde{E}(\vec{R})$. The tilde denotes that the quantity is estimated. $P$ is a set of parameters introduced by the ML model. 
The idea of the ML approaches is to sample configurations $\vec{R}$ on a PES, and then fit the parameters $P$ in the function $f_P$ to obtain an accurate description of $\tilde{E}(\vec{R})$. The configurations $\vec{R}$ can be transformed using a function $h$ before being fed into the ML model. An example would be a distance transformation (distance matrix). The parameters $P$ of the ML model are fitted such that the loss $L_P(E(\vec{R}),\tilde{E}(\vec{R})) = \text{argmin}_P(g_c(E(\vec{R}),\tilde{E}(\vec{R})))$ becomes minimal, where $g_c$ is an arbitrary convex function. Examples are mean squared errors and mean absolute errors. $\tilde{E}(\vec{R})$ can subsequently be used to propagate an MD trajectory.\cite{Behler2015} 

Different target outputs can be chosen. Instead of learning the complete potential $\tilde{E}(\vec{R})$, it is possible to learn only the difference between two levels of theory (e.g. semi-empirical and DFT),\cite{Brunk2015}
\begin{equation}
    \Delta E(\vec{R}) = E^{\text{expensive}}(\vec{R}) - E^{\text{cheap}}(\vec{R}).
    \label{eq:delta_definition}
\end{equation}
The function $f$ thus becomes a mapping of $f_P:h(\vec{R})\rightarrow \Delta \tilde{E}(\vec{R})$.
Alternatively, it is possible to directly learn the gradient $f_P:h(\vec{R})\rightarrow \frac{\partial E(\vec{R})}{\partial \vec{R}}$\cite{Botu2015}. An obstacle when learning gradients is that $-\int_{-\infty}^{\infty}\text{d}\vec{R} \frac{\partial E(\vec{R})}{\partial \vec{R}}=E$ must be ensured. This can for example be achieved with the sGDML method.\cite{chmiela2017,Chmiela2018}
As a third option, information about the gradients can be included in the training process. In other words, an additional loss term $L_P(E,\tilde{E}) = \text{argmin}_P(g_c(E,\tilde{E})+g_c(\frac{\partial E}{\partial \vec{R}},\frac{\partial \tilde{E}}{\partial \vec{R}}))$ is added to account for the gradients.
The concepts described above are independent of the ML method used and the transformation $h$ applied. It is essential that $h$ does not remove information relevant for the target output.

\subsubsection{High-Dimensional Neural Network Potentials (HDNNP)}
Feed-forward neural networks (FNN) consist of a number of sequentially chained (non)-linear functions $\sigma^l$, weight matrices $W^l$ and biases $b^l$, where $l$ is the number of layers.
A FNN can be defined recursively as,\cite{Alpaydin2014}
\begin{equation}
    x^{l+1} = \sigma^l (W^lx^{l}+b^l),
\label{eq:FNN}
\end{equation}
where $x$ is the input vector. The weight matrices and biases correspond to the parameters $P$. In addition, there are user-defined hyperparameters such as the number of layers $l$, the number of neurons per layer, and the type of functions $\sigma^l$.

FNN are the method of choice in cases, where a memory intensive data set is required and complicated transformations are necessary in order to learn the underlying relation between the input $z$ and the output $\tilde{E}$. Moreover, they provide more flexibility compared to Kernel ridge regression and Gaussian processes. 
It has been observed in Kaggle contests\cite{Klambauer2017} that two hidden layers usually perform best if no special activation functions are used. In addition, FNN with a lower number of parameters tend to generalize better and require smaller training sets.\cite{MorphNet} 

The FNN architecture does not contain any information about the parity of the system. In chemistry, this is an issue because the rotation and translation of the system or swapping of atoms of the same type need to leave the energy of the system unchanged. It is possible to augment the training data set in order to force the FNN to learn the correct symmetry. However, this increases the training set by multiple orders of magnitude. Behler \textit{et al.}\cite{Behler2011} identified this issue and developed a FNN architecture and an input-coordinate transformation $h$, which guarantees the correct parity behaviour, the so-called high-dimensional neural network potentials (HDNNP). Instead of using only one FNN, each element type in the system is described by its own FNN. Each FNN is called an NNP because it describes the atomic energy of the respective atom type. The total energy of the system $E$ is thus separated, i.e., $E = \sum_i^{N_\text{H}} E_i^{\text{H}}+\sum_i^{N_\text{C}}E_i^{\text{C}}+...$, where $N_\text{H}$ is the number of hydrogen atoms, $N_\text{C}$ the number of carbon atoms, and so on. Note that this separation ansatz is an approximation, it has no quantum-mechanical foundation. It guarantees that the total energy of a system is invariant with respect to swapping the positions of two atoms of the same type. However, rotation and translation still change the total energy. To address this issue, Behler \textit{et al.} used the concept of symmetry functions.\cite{Behler2011} The Cartesian coordinates are transformed using a set of symmetry functions, which are invariant with respect to rotation and translation,\cite{Behler2011,Smith2017}
\begin{equation}
    S_{i}^{t,\text{Rad}}(\vec{R}) = \sum_j e^{-\eta_{\text{Rad}} \cdot (R_{ij}-\mu_{\text{Rad}})^2}
\label{eq:radial}
\end{equation}
and
\begin{equation}
S_{i}^{t,\text{Ang}}(\vec{R}) = 2^{\zeta-1} \sum_{j\neq k\neq i} (1-\text{cos}(\theta_{ijk}-\theta_S))^\zeta\cdot e^{-(\frac{R_{ij}-R_{ik}}{2}-\mu_{\text{Ang}})^2\cdot \eta_{\text{Ang}}},
\label{eq:angle}
\end{equation}
where $R_{ij}\equiv |\vec{R}_i-\vec{R}_j|$, and $\theta_{ijk}$ is the angle spanned by atoms $i$, $j$ and $k$. The additional parameters $\eta_{\text{Rad}}$, $\mu_{\text{Rad}}$, $\mu_{\text{Ang}}$, $\zeta$, $\theta_S$ and $\eta_{\text{Ang}}$ need to be set manually.  The superscript $t$ (for type) indicates that the summation only runs over a specific element type. ``Rad" stands for radial and ``Ang" for angular. Symmetry functions describe the chemical environment of each atom with radial (Eq.~(\ref{eq:radial})) and angular features (Eq.~(\ref{eq:angle})). The summation operator enforces correct parity behaviour.\cite{summation}

Applying symmetry functions to periodic boundary conditions requires a cutoff. For this, a cutoff function $f_C$ can be added with the cutoff parameters $R_{\text{sym,Rad}}$ (for radial symmetry functions) and $R_{\text{sym,Ang}}$ (for angular symmetry functions), i.e.,
\begin{equation}
    S_{i}^{t,\text{Rad}}(\vec{R}) = \sum_j e^{-\eta_{\text{Rad}} \cdot (R_{ij}-\mu_{\text{Rad}})^2}\cdot f_C(R_{ij};R_{\text{sym,Rad}})
\label{eq:radialcut}
\end{equation}
and
\begin{equation}
S_{i}^{t,\text{Ang}}(\vec{R}) = 2^{\zeta-1} \sum_{j\neq k\neq i}^{N} (1-\text{cos}(\theta_{ijk}-\theta_S))^\zeta\cdot e^{-(\frac{R_{ij}-R_{ik}}{2}-\mu_{\text{Ang}})^2\cdot \eta_{\text{Ang}}}\cdot f_C(R_{ij};R_{\text{sym,Ang}})\cdot f_C(R_{ik};R_{\text{sym,Ang}}) ,
\label{eq:anglecut}
\end{equation}
with the cutoff function $f_C$ defined as,
\begin{equation}
    f_C(R_{ij};R_{\text{sym}}) = \begin{cases}
    0.5\cdot(1+\text{cos}(\frac{R_{ij}\cdot \pi}{R_{\text{sym}}}))  & \text{for } R_{ij} \leq R_{\text{sym}}  \\
    0 & \text{else.}
    \end{cases}
\end{equation}
Note that the symmetry functions $S$ take over the role of the function $h$ introduced in the previous section.
To mimic long-range interactions, a second ML model can be trained on QM-derived partial charges and multipole moments to be used in an Ewald summation scheme.
For covalently bonded or metallic systems, the cutoff can be chosen relatively small because the contributions of the long-range interactions are minor.
However, as discussed in the Introduction, much larger cutoffs are required for condensed-phase systems.

HDNNP present a powerful technique as these models can learn the complete Born-Oppenheimer approximated PES (given an appropriate training set), i.e.,
\begin{equation}
\underbrace{E_{\text{QM}}(\vec{R})}_{\text{Learned by HDNNP}} = \frac{\braket{\psi(\vec{r})|\hat{H}_\text{QM}\psi(\vec{r})}}{\braket{\psi(\vec{r})|\psi(\vec{r})}}.
\label{eq:learnedfull}
\end{equation}
Applying $\Delta$-learning and using the definition of $\Delta E$ in Eq.~(\ref{eq:delta_definition}), we can write
\begin{equation}
E_{\text{QM}}^{\text{cheap}}(\vec{R})+\underbrace{\Delta E(\vec{R})}_{\text{Learned by HDNNP}} = \frac{\braket{\psi(\vec{r})|\hat{H}_\text{QM}\psi(\vec{r})}}{\braket{\psi(\vec{r})|\psi(\vec{r})}},
\label{eq:learneddelta}
\end{equation}

However, HDNNP are limited in the following ways: 
First, the number of angle terms of the symmetry functions scale exponentially with the number of element types as all possible combinations must be considered. For example, a system consisting of hydrogen and carbon atoms involves five unique angle combinations. 
Introducing an additional element type (e.g. oxygen) adds five more unique combinations. 
This issue can be resolved partially by introducing weighted symmetry functions. Instead of introducing a new element type, the symmetry functions are weighted with respect to their atomic number and/or charge.\cite{Gastegger2018}
Other issues with HDNNP are that the full correct permuting behaviour is only partially recovered,\cite{Jun2019} and that the descriptors used are non-unique.\cite{Grisafi2019} However, the latter issue is in our opinion more of theoretical nature, and should not limit the usage of HDNNPs in practical applications. Lastly, one has to keep in mind that HDNNPs do not resolve the inherent general limitations of ML approaches. For example, the requirement for large training sets, danger of overfitting, and limited extrapolation capabilities apply also to HDNNPs.

\subsection{(QM)ML/MM MD Simulations Using HDNNP}
In the original work, HDNNP are used to learn the complete Born-Oppenheimer approximated PES.
In this work, we propose to combine the HDNNP with classical force fields in a QM/MM-type approach to reduce the complexity of learning. Hence, instead of Eq.~\ref{eq:learnedfull}, our target is
\begin{equation}
    \underbrace{E_{\text{QM}}(\vec{R}_{\text{QM}})+E^\text{el}_\text{QM--MM}(\vec{R})}_{\text{Learned by HDNNP}}+E_{\text{MM}}(\vec{R}_\text{MM})+E^{\text{vdW,SR}}_{\text{QM--MM}}(\vec{R}) = E(\vec{R}),
\end{equation}
or in the case that $\Delta$-learning is applied it is,
\begin{align} \nonumber
    E_{\text{QM,cheap}}(\vec{R}_{\text{QM}}) &+ E^\text{el}_\text{QM--MM,cheap}(\vec{R}) + \underbrace{\Delta E_{\text{QM}}(\vec{R}_{\text{QM}}) + \Delta E^\text{el}_\text{QM--MM}(\vec{R})}_{\text{Learned by HDNNP}} \\ 
    &+ E_{\text{MM}}(\vec{R}_\text{MM}) + E^{\text{vdW,SR}}_{\text{QM--MM}}(\vec{R}) = E(\vec{R}).
\end{align}
$E_\text{MM}(\vec{R}_{\text{MM}})$ and $E^{\text{vdW,SR}}_\text{QM--MM}(\vec{R})$ are provided by the classical force field. Thus, it is expected that the complexity of the model can be reduced drastically. Furthermore, long-range interactions are now included via point charges, allowing to use a larger cutoffs in the construction of the HDNNP training set.
For this approach, it is necessary to incorporate the MM environment into the descriptor of the ML model.
The MM environment can be encoded in the HDNNP by introducing the MM particles as an additional element type, which is weighted with the respective partial charge.
For the QM particles, we obtain
\begin{equation}
    S_{i}^{t,\text{Rad}}(\vec{R}_\text{QM}) = \sum_j^{N(t)_{\text{QM}}} e^{-\eta_{\text{Rad}} \cdot (R_{ij}-\mu_{\text{Rad}})^2}\cdot f_C(R_{ij};R_{\text{sym},\text{Rad}}^{\text{QM-QM}}))
\label{eq:radialcut2}
\end{equation}
and
\begin{align} \nonumber
S_{i}^{t,\text{Ang}}(\vec{R}_\text{QM}) &= 2^{\zeta-1} \sum_{j\neq k\neq i}^{N(t)_{\text{QM}}} (1-\text{cos}(\theta_{ijk}-\theta_S))^\zeta\cdot e^{-(\frac{R_{ij}-R_{ik}}{2}-\mu_{\text{Ang}})^2\cdot \eta_{\text{Ang}}} \\
& \cdot f_C(R_{ij};R_{\text{sym,Ang}})\cdot f_C(R_{ik};R_{\text{sym,Ang}}).
\label{eq:anglecut2}
\end{align}
MM particles are introduced as a new element type by
\begin{equation}
    S_{i}^{t,\text{Rad}}(\vec{R}) = \sum_j^{N_{\text{MM}}} e^{-\eta_{\text{Rad}} \cdot (R_{ij}-\mu_{\text{Rad}})^2}\cdot f_C(r_{ij};R_{\text{sym},\text{Rad}}^{\text{QM-MM}})\cdot Z(j) ,
\label{eq:radialcutMM}
\end{equation}
\begin{align} \nonumber
S_{i}^{t,\text{Ang}}(\vec{R}) &= 2^{\zeta-1} \sum_{j\neq i }^{N(t)_{\text{QM}}}\sum_{k}^{N_{\text{MM}}} (1-\text{cos}(\theta_{ijk}-\theta_S))^\zeta\cdot e^{-(\frac{R_{ij}-R_{ik}}{2}-\mu_{\text{Ang}})^2\cdot \eta_{\text{Ang}}} \\
& \cdot f_C(R_{ij};R_{\text{sym,Ang}})\cdot f_C(R_{ik};R_{\text{sym,Ang}})\cdot Z(k) ,
\label{eq:anglecut_1MM}
\end{align}
and
\begin{align} \nonumber
S_{i}^{t,\text{Ang}}(\vec{R}) &= 2^{\zeta-1} \sum_{j\neq k}^{N_{\text{MM}}} (1-\text{cos}(\theta_{ijk}-\theta_S))^\zeta\cdot e^{-(\frac{R_{ij}-R_{ik}}{2}-\mu_{\text{Ang}})^2\cdot \eta_{\text{Ang}}} \\
& \cdot f_C(R_{ij};R_{\text{sym,Ang}})\cdot f_C(R_{ik};R_{\text{sym,Ang}})\cdot Z(j) \cdot Z(k).
\label{eq:anglecut_2MM}
\end{align}
In Eqs. (\ref{eq:radialcutMM}-\ref{eq:anglecut_2MM}), the summation operator iterates over all MM particles $N_\text{MM}$ and QM particles $N(t)_\text{QM}$ of type $t$. The function $Z(i)$ returns the partial charge of the MM particle. For SPC/E water, we have $Z(\text{O})=-0.8476 $ and $Z(\text{H})=0.4238$. 
$R_{\text{sym},\text{Rad}}^{\text{QM-QM}}$ and $R_{\text{sym},\text{Rad}}^{\text{QM-MM}}$ are cutoff parameters for the radial symmetry functions, the first one influencing the QM-QM interaction and the second one targeting the inclusion of the MM particles.

This approach requires that the data set is generated within the same QM/MM approach. No partitioning of the QM zone or MM zone is performed. The model $f_P$ is then trained on the generated data points. The learned potential $\tilde{E}_{\text{QM--MM}}^{\text{el}}(\vec{R})+\tilde{E}_{\text{QM}}(\vec{R}_{\text{QM}})$ in combination with the classical force field can subsequently be used to propagate the total system in time.

\section{Methods}
\subsection{Systems}
Two types of systems were investigated: (i) single molecules in water, including benzene, uracil and retionic acid, and (ii) chemical reactions in water (constrained close to the transition state), namely the S$_N$2 reaction of CH$_3$Cl with Cl$^-$ and the reaction of S-adenosylmethionate (SAM) with cytosine. The systems were chosen to represent different system sizes, use cases, and difficulties. 
The systems were either treated in a \textit{static} or \textit{dynamic} manner. Static refers to a retrospective evaluation of sampled configurations, whereas dynamic indicates that an actual MD simulation was performed using the gradients provided by the ML model.

\subsection{General Computational Details}
All MD simulations were performed using the GROMOS software package\cite{schmid2012,meier2012} interfaced to DFTB+/19.2\cite{DFTB} and ORCA/4.2.0\cite{Neese2018}.
The functionals used in this study are the gradient corrected Becke-Perdew functional (BP86), Head-Gordon's range-separated functional $\omega$B97X-D3\,\cite{Chaia2008}, and Grimme's double hybrid functional B2-PLYP\cite{Grimme2006}. The basis set used is Ahlrichs and Weigend's def2-TZVP\cite{Weigend2005}.

All QM calculations used the resolution of identity\cite{Feyeresein1993}, Weigend's auxiliary basis\cite{Weigend2006} functions, and Grimme's dispersion correction with Becke-Johnson damping\cite{Grimme2010,Grimme2011} (except for $\omega$B97X-D3). 
DFT calculations, which require the computation of a Hartree-Fock\cite{Hartree} reference wave function, were further accelerated using the so-called COSX\cite{Kossmann2009} approximation. The ORCA computations used TightSCF convergence criteria and the integration Grid5. The grid was changed to Grid6 in the final iteration. 
Otherwise, standard parameters were used. 
DFTB computations were performed with Grimme's D3 dispersion correction with Becke-Johnson damping.\cite{Grimme2010,Grimme2011} 

All point charges within the cutoff radius $R_c$ were included in an electrostatic embedding scheme in the QM computation. Selected solvent atoms beyond the cutoff were included to avoid bond-breaking and creation of artificial charges. The convergence criteria was set to $10^{-8}$\,eV for all systems. Structures, which did not converge in the maximum number of steps, were discarded. MD simulations were performed with a convergence criteria of $10^{-6}$\,eV and a Broyden mixer\cite{Broyden1965} with a mixing parameter of $\alpha=0.3$ to ensure numerical stability. 

Newton's equations of motion were integrated with a time step of 0.5\,fs. The temperature was kept constant in the MD simulations with the Nos\'e-Hover chain\cite{Nose1984,Hoover1985} thermostat with a coupling constant of 0.1\,ps and two baths (one coupled to the internal motion and the rotation of the solute, and the other one to the solvent). A weak-coupling\cite{berendsen1984} barostat was used for constant pressure simulations with a coupling constant of 0.5\,ps and a isothermal compressibility of $4.575\cdot10^{-4}$\,(kJ\,mol$^{-1}$\,nm$^{-3}$)$^{-1}$. Long-range electrostatic interactions beyond the cutoff of 1.4\,nm were included using a reaction-field method.\cite{Tironi1995} Note that the reaction field acts only on the MM particles. The water model used in this study was SPC/E\cite{Berendsen1987}. 
All bonds between MM particles were constrained using the SHAKE algorithm\cite{ryckaert1977} and a relative tolerance of 10$^{-4}$. The motion of the center of mass was removed every 1000 steps.
A topology file of the QM solute is required as input in GROMOS. The Lennard-Jones parameters are needed to evaluate $E_\text{QM--MM}^{\text{vdW,SR}}$, since this interaction between the QM zone and MM zone is treated on the classical level. All other parameters are discarded for the QM solute. The topology files were obtained from the ATB\cite{Malde2011} server.

The neural networks were implemented using Tensorflow/keras.\cite{tensorflow2015-whitepaper} We exported the computational graphs trained in Python to the C++ GROMOS code.

\subsection{Initial Structures}
Initial structures of the individual solutes were generated using the ATB\cite{Malde2011} server. Initial structures for the chemical reactions were generated by placing the reactants manually close to the transition state. During sampling, a biasing potential was applied of the form,
\begin{equation}
    V^{\text{bias}} = \frac{1}{2}k(R_{ij}+d\cdot R_{kl}-R_0)^2 ,
\end{equation}
where $k$ is the force constant, and $R_{ij}$, $R_{kl}$ are the distances between the atoms $i$ and $j$, and $k$ and $l$, respectively. $R_0$ is the ideal distance, and $d\in \{-1,+1\}$ determines, whether the distances are subtracted or added. For both systems, we set $d=-1$ and $R_0=0.00$.
For the CH$_{3}$Cl/Cl$^{-}$ system, the indexing is shown in the following,
\begin{equation}
\text{Cl}_{j}\text{C}_{i,k}\text{H}_3 + \text{Cl}_{l}^{-} \rightarrow \text{ClCH}_3 + \text{Cl}^{-}.
\end{equation}
The structure was minimized using a force constant for the biasing potential $k$~=~2'000~kJ~mol$^{-1}$~nm$^{-2}$.

For the SAM/cytosine system, atom $i$ was the carbon atom participating in the reaction, atom $j$ the sulfur atom, atom $k$ was the same carbon atom as atom $i$, and atom $l$ corresponds to the carbon atom in $\alpha$-position to the amine group.

For all systems, the solute(s) was solvated in a periodic box of SPC/E water using the GROMOS++ package of programs,\cite{Eichenberger2011} with a minimum solute-wall distance of 1.6\,nm and a minimum solute-solvent distance of 0.25\,nm. The size of the simulation boxes were between 3.3\,nm and 5.0\,nm. All systems were initially relaxed at 0\,K using a gradient descent algorithm. A configuration was considered to be converged when the predicted change in energy was less than 0.1\,kJ\,mol$^{-1}$ averaged over all particles. 
The thermostat temperature was set to 298\,K or 400\,K with a coupling constant of 0.1\,ps.

\subsection{Sampling of Configurations}
All systems were simulated for 20'000 steps at $T=400$\,K, $p=1$\,bar. The first 10'000 steps were discarded as equilibration. The remaining 10'000 frames were split into 70/20/10 training/validation/test sets. Thus, the last 1'000 steps were considered as the external test set.
The performed splitting mimics the envisioned use case and gives a validation and test set, which do not resemble the training set as much as if a random partitioning would have been performed.

For benzene in water, a cutoff for the partial charges in the QM/MM Hamiltonian of $R_c=0.6$\,nm was used. For all other systems, the cutoff for the partial charges was set to $R_c=1.4$\,nm. For CH$_3$Cl/Cl$^{-1}$ and SAM, a biasing potential with $k$~=~2'000~kJ~mol$^{-1}$~nm$^{-1}$ was employed. Note that for the training of the HDNNP, the biasing potential is removed.
Commonly used sampling strategies such as normal-mode sampling are difficult to employ for these systems as they do not account for the solvent environment. 

\subsection{MD Simulations Using the Fitted ML Models}
For the large systems, (QM)ML/MM MD simulations were performed using the fitted ML models. The time step was set for both system to 0.5\,fs, the temperature was set to $T=298$\,K, the pressure was set to 1\,bar. For the SAM/cytosine system, the biasing force constant was set to $k$~=~2'000~kJ~mol$^{-1}$~nm$^{-2}$.

\subsection{Neural Networks}
We implemented the HDNNP in Tensorflow/keras with a float64 precision. Each HDNNP used at most two hidden layer and thus followed the recommendation in Ref.~\citenum{Klambauer2017}. The activation function mila\cite{mila} was used with a coefficient $\beta=-0.25$, because it is continuously-differentiable and is known to outperform commonly used activation functions such as tanh. The ML models were trained using the Adam optimizer\cite{Kingma2015} with varying learning rates. The number of neurons per hidden layer ranged for each NNP from 10 to 80 neurons. As a loss function, we used
\begin{equation}
    L = \frac{1}{N}\cdot \sum_i^{N} (E_i - \tilde{E}_i)^2
\end{equation}
and 
\begin{equation}
    L' = L+\frac{\omega_0}{3N_\text{QM}}\cdot \sum_i^{N_\text{QM}}\sum_\alpha^3 (-F_{i\alpha} + \tilde{F}_{i\alpha})^2+\frac{\omega_1}{3N_\text{MM}}\cdot \sum_i^{N_\text{MM}}\sum_\alpha^3 (-F_{i\alpha} + \tilde{F}_{i\alpha})^2 ,
    \label{eq:gradient_corrected}
\end{equation}
where $N_\text{MM}$ is the number of MM particles, $N_\text{QM}$ the number of QM particles, and $\omega_0$ and $\omega_1$  are weight parameters for the gradient contribution. If not noted otherwise, $\omega_0=1$ and $\omega_1=1$.
Note that $L$ is a non-gradient corrected loss function and $L'$ is a gradient corrected loss function.
All models were initially trained for 50 steps with a learning rate of $3.5\cdot 10^{-4}$. The SAM/cytosine system was trained for 20 epochs. We monitored the validation loss during the training process. After training, the model with the lowest loss on the validation set was recovered.

\subsection{Symmetry Functions}
The symmetry functions were extracted from Refs.~\citenum{Smith2017,Stevenson2019}. In contrast to Smith \textit{et al.}, we did not apply any prefactors. In addition, we allowed for a longer cutoff and extended the spacing of the radial functions. Table\,\ref{tab:sym} lists the parameters for the symmetry functions in Eqs.~(\ref{eq:radial}) - (\ref{eq:anglecut_2MM}), which encode the environment. The cutoffs for the radial symmetry functions were varied in this study and are provided in the Results and Discussion section for all models.

\begin{table}[H]
    \centering
    \begin{tabular}{|c|c|c|c|c|}
    \hline
    \textbf{Parameter} & \textbf{Type} & \textbf{From} & \textbf{To} & \textbf{Step size} \\\hline \hline
    $\mu_{\text{Rad}}$   & Radial  & 0.09 & 1.40 & 0.026875  \\\hline
    $\eta_{\text{Rad}}$    &Radial&160&160 &0 \\\hline
    $\mu_{\text{Ang}}$   & Angular  & 0.09 & 0.285 & 0.065 \\\hline
    $R_{\text{sym,Ang}}$     &Angular&0.35&0.35 &0 \\\hline
    $\eta_{\text{Ang}}$  &Angular&80&80 &0 \\\hline
    $\zeta$     &Angular&32&32 &0 \\\hline
    $\theta_S$  &Angular&0.19634954&2.9452431& 0.19634954\\\hline
    \end{tabular}
    \caption{Parameters of the symmetry functions used in this work, see Eqs.~(\ref{eq:radial}) - (\ref{eq:anglecut_2MM}). The symmetry functions were taken from Ref.~\citenum{Stevenson2019} and adapted to units nm and nm$^{-1}$. The cutoffs for the radial symmetry functions for the different models are provided in the Results and Discussion section.}
    \label{tab:sym}
\end{table}
For the MM particles, we used either the same symmetry functions as for the QM subsystems (two-body terms and three-body terms), or omitted the three-body terms (only two-body terms). All symmetry functions that include MM particles were weighted with the corresponding partial charge. Note that symmetry functions could be varied to optimize the results. However, Smith \textit{et al.}\cite{Smith2017,Stevenson2019} have shown that their set of symmetry functions covers important local interactions and torsional space in organic compounds, while still being computationally efficient and less prone to overfitting.

In the following, we will refer to three cutoffs:
$R_c$ is the cutoff chosen in the QM/MM calculation to generate the data sets. The summation in Eq.\,(\ref{eq:QMMMH}) goes over all partial charges within $R_c$. Similarly, the pairwise van der Waals interactions in Eq.~(\ref{eq:SR}) are calculated within $R_c$.
$R_{\text{sym,Rad}}^{\text{QM-QM}}$ is the cutoff used for the radial symmetry functions encoding the QM environment (see Eq.\,(\ref{eq:radialcut2})), whereas $R_{\text{sym,Rad}}^{\text{QM-MM}}$ is the cutoff used for the radial symmetry functions encoding the MM environment (see Eq.\,(\ref{eq:radialcutMM})). $R_{\text{sym,Rad}}^{\text{QM-QM}}$ and  $R_{\text{sym,Rad}}^{\text{QM-MM}}$ correspond to the cutoffs used in the HDNNP.

\section{Results and Discussion}

\subsection{Assessing the Effect of Different Parameters on the Prediction Accuracy}
The loss function of neural networks has multiple local minima. The training procedure usually finds a configuration of weight parameters close to a local minimum, but not necessarily close to the global minimum. To assess the impact of gradients correction, many-body terms, and the cutoff size, we chose test systems with a conformationally rigid solute in water where the conformational space can be sampled completely. Three different solutes were investigated, which differ in the polarity: (i) benzene (apolar), uracil (polar), and the transition state of CH$_3$Cl/Cl$^{-}$ (charged). The trajectories were split into 70/20/10 training/validation/test sets (see Methods). As discussed in the Theory section, the ML models learn the quantity $E_\text{QM}(\vec{R}_{\text{QM}})+E^{\text{el}}_{\text{QM--MM}}(\vec{R})$. The performance of the ML models are thus assessed based on the error on this quantity (energy), and on the respective derivatives (forces), i.e. the gradients on the QM particles
\begin{equation}
    \frac{\partial (E_\text{QM}(\vec{R}_{\text{QM}})+E^{\text{el}}_{\text{QM--MM}}(\vec{R}))}{\partial \vec{R}_{\text{QM}}} = -\vec{F}_\text{QM}\, 
\end{equation}
and the gradients on the MM particles resulting from the QM zone,
\begin{equation}
    \frac{\partial ( E_\text{QM}(\vec{R}_{\text{QM}})+E^{\text{el}}_{\text{QM--MM}}(\vec{R}))}{\partial \vec{R}_{\text{MM}}} = -\vec{F}_\text{MM}. 
\end{equation}

\subsubsection{Gradient Correction and Many-Body Terms}
The first test system consists of benzene in water. Benzene is an apolar solute. Thus, long range interactions do not contribute significantly to the potential energy compared to polar solutes. Point charges further away than 0.6\,nm from the solute atom were not included in the evaluation of Eq.\,(\ref{eq:QMMMH}). The test system is suitable to study the effect of including gradient correction and many-body terms (Eqs. (\ref{eq:radialcutMM}) - (\ref{eq:anglecut_2MM})) in the description of the PES for the MM part. Note that the descriptor always contains many-body information for the QM particles (Eq. (\ref{eq:anglecut})).
Table \ref{tab:ID} summarizes the settings of four ML models used for the comparison. All models use the parameters proposed by Ref.~\citenum{Smith2017}. However, we account for the interactions of the QM subsystem with the MM particles by introducing them as a new element type (see Theory section).
Model 1 (M1) is the baseline model, including only two-body information and no gradient correction. M2 includes three-body terms for the MM point charges. M3 only includes gradient information. M4 additionally weights the gradients of the MM particles in the loss function. Note that the cutoffs $R_{\text{sym,Rad}}^{\text{QM-QM}}$ and $R_{\text{sym,Rad}}^{\text{QM-MM}}$ for M1-M4 are larger than $R_c$. This means that the models ``know'' about all particles contributing to the potential energy.

\begin{table}[H]
    \centering
    \scalebox{0.8}{
    \begin{tabular}{|c|c|cc|ccc|}
    \hline
    \textbf{Settings ID}  & \textbf{Three-body terms} & \multicolumn{2}{c|}{\textbf{Gradient correction}} & \multicolumn{3}{c|}{\textbf{Cutoff}}\\ 
    & \textbf{for MM?} & $\omega_0$ & $\omega_1$ & $R_{\text{sym,Rad}}^{\text{QM-QM}}$ [nm] &$R_{\text{sym,Rad}}^{\text{QM-MM}}$ [nm] & $R_c$ [nm] \\ \hline \hline
    M1    &No&0&0&1.0&1.0&0.6\\\hline
    M2    &Yes&0&0&1.0&1.0&0.6\\\hline
    M3    &No&1&1&1.0&1.0&0.6\\\hline
    M4    &No&1&1-200&1.0&1.0&0.6\\\hline
    \end{tabular}}
    \caption{Settings for the ML models used to test the influence of including gradient correction or three-body terms for the MM point charges. $\omega_0$ and $\omega_1$ are weight parameters for the gradient contribution (Eq. (\ref{eq:gradient_corrected})). $R_c$ is the cutoff radius for the MM partial charges in the QM/MM calculation to generate the data set, $R_{\text{sym,Rad}}^{\text{QM-QM}}$ is the cutoff for the QM-QM symmetry functions, and $R_{\text{sym,Rad}}^{\text{QM-MM}}$ the cutoff for the QM-MM symmetry functions. The test system is benzene in water.}
    \label{tab:ID} 
\end{table}

\begin{figure}[H]
    \centering
    \includegraphics[width=\textwidth]{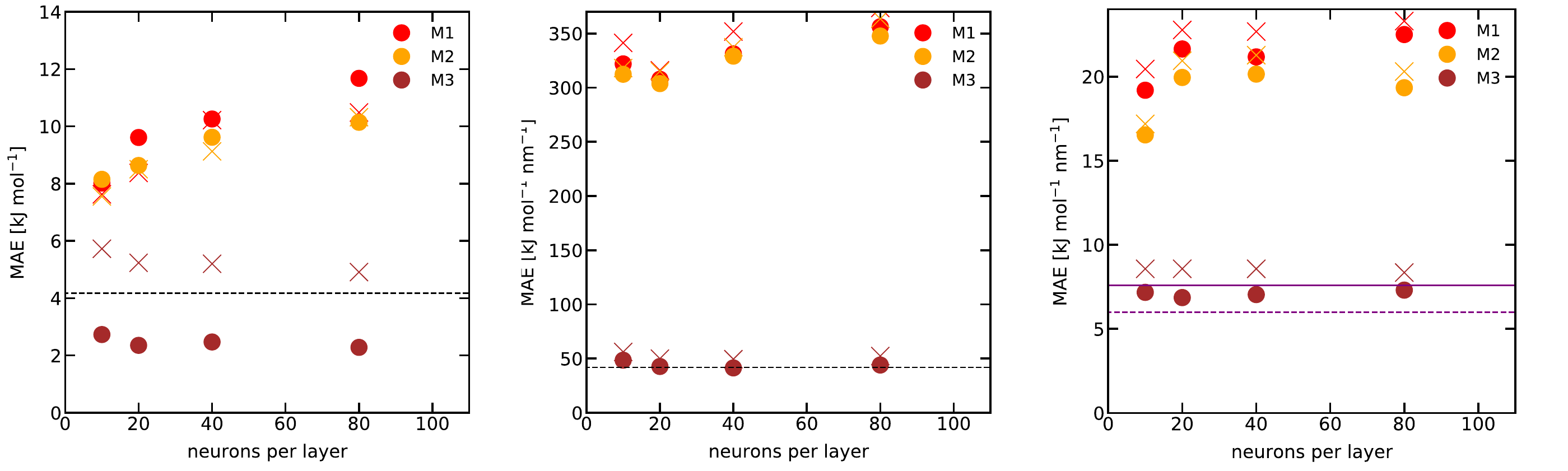}
    \caption{Performance of the settings M1-M3 for benzene in water as a function of the number of neurons in each hidden layer. The reference data points were computed with B2-PLYP/def2-TZVP. The performance is reported for the validation set (filled circles) and the test set (crosses) as the mean absolute error (MAE). The black dashed line indicates chemical accuracy (i.e. 4.18 kJ mol$^{-1}$ for the energies, and 41.8 kJ mol$^{-1}$ nm$^{-1}$ for the gradients). (Left): MAE of the predicted energies. (Middle): MAE of the predicted QM gradients. (Right): MAE of the predicted MM gradients. For comparison, the purple lines denote the performance of DFTB on the validation set (dashed line) and test set (solid line).} 
    \label{fig:neurons}
\end{figure}
We first analyzed the model performance as a function of the number of neurons in the hidden layers for M1 and M2. The accuracy on the energies of the validation/test set is far above chemical accuracy for both models, and also the performance on the QM gradients (mid panel) and the MM gradients (right panel) is poor. Thus, including three-body terms in the descriptor for the MM particles (M2) does not improve the accuracy. Interestingly, the performance decreases even further with an increasing number of neurons in the hidden layers. This is in line with the observation in ML that large numbers of weight parameters should be avoided, when regularization techniques are not applied. 

M3 includes gradient information of the QM and MM particles in the training procedure, and thus regularizes the weights of the models. The accuracy on the energies of the validation set is now clearly below chemical accuracy, and also the performance on the gradients is significantly improved. We observe that the performance is stabilized, meaning that a larger number of neurons per layer slightly improves the performance instead of worsen it.

\begin{figure}[H]
    \centering
    \includegraphics[width=\textwidth]{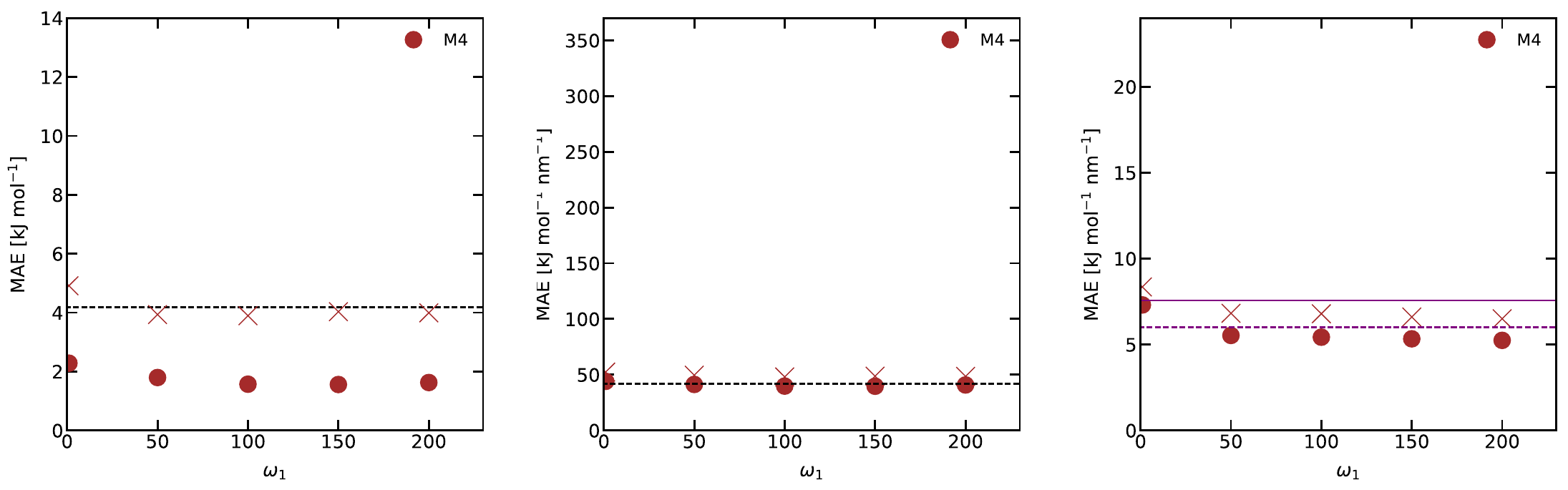}
    \caption{Performance of the setting M4 for benzene in water as a function of the weight $\omega_1$ (see Eq.\,(\ref{eq:gradient_corrected})). The reference data points were computed with B2-PLYP/def2-TZVP. All models used 80 neurons per layer. The performance is reported for the validation set (filled circles) and the test set (crosses) as the mean absolute error (MAE). The black dashed line indicates chemical accuracy (i.e. 4.18 kJ mol$^{-1}$ for the energies, and 41.8 kJ mol$^{-1}$ nm$^{-1}$ for the gradients). (Left): MAE of the predicted energies. (Middle): MAE of the predicted QM gradients. (Right): MAE of the predicted MM gradients. For comparison, the purple lines denote the performance of DFTB on the validation set (dashed line) and test set (solid line).}
    \label{fig:weights} 
\end{figure}

The influence of regularizing the models via gradients can be further highlighted by varying the parameter $\omega_1$. In Eq.\,(\ref{eq:gradient_corrected}), $\omega_1$ scales the contribution of the error on the MM gradients. The gradients on the MM particles are much smaller than the QM gradients, since the quantity only incorporates the electrostatic interaction with the QM zone, and not the interaction of the MM particles with themselves. The latter interactions as well as the short-range van der Waals interactions between the QM and MM particles are modulated with the classical force field. Thus, the contribution of the MM particles to the loss function is substantially less than the contribution of the energy and forces of the QM particles if $\omega_1=1$ is used. The performance on the training set measured by the loss function is thus significantly increased at the cost of a worse description of the interaction with the MM particles. This behaviour is clearly non-physical and leads to a HDNNP, which performs worse on the validation (and test) set. Scaling the MM gradients via $\omega_1$ solves this issue. Figure~\ref{fig:weights} shows the performance of M4 on the energies and gradients as a function of the parameter $\omega_1$. By increasing $\omega_1$, the prediction accuracy can be further improved. 

\subsubsection{$\Delta$-Learning}
\begin{figure}[H]
    \centering
    \includegraphics[width=\textwidth]{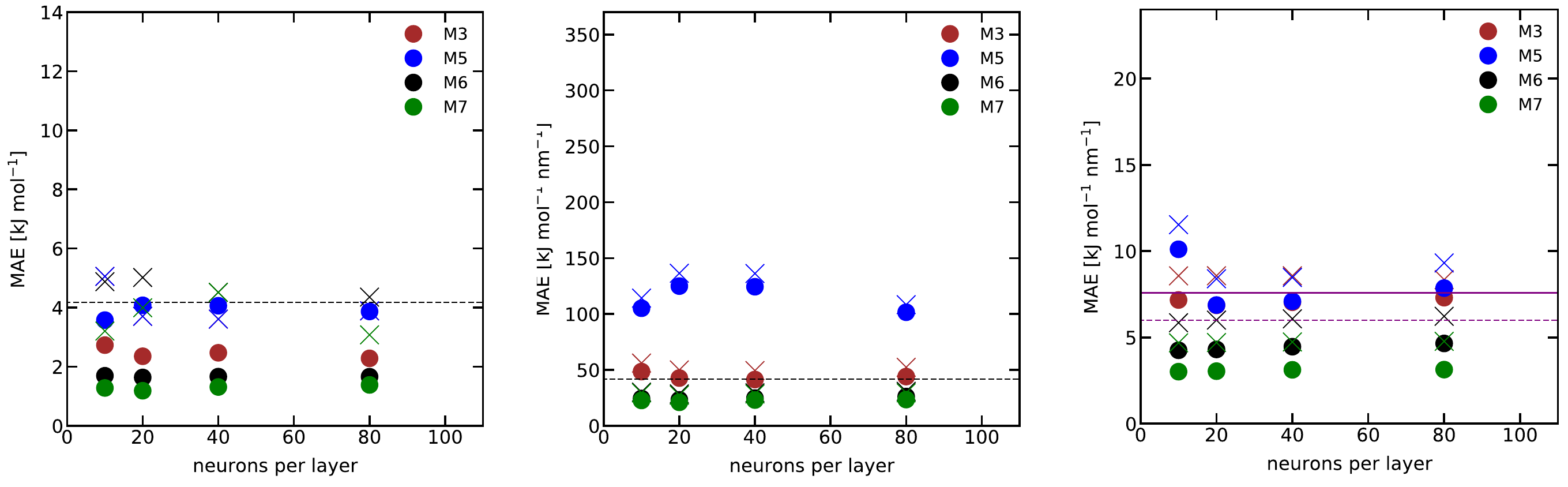}
    \caption{Performance of the settings M3 (no $\Delta$-learning) and M5-M7 ($\Delta$-learning) for benzene in water as a function of the number of neurons in each hidden layer. The reference data points were computed with B2-PLYP/def2-TZVP. The performance is reported for the validation set (filled circles) and the test set (crosses) as the mean absolute error (MAE). The black dashed line indicates chemical accuracy (i.e. 4.18 kJ mol$^{-1}$ for the energies, and 41.8 kJ mol$^{-1}$ nm$^{-1}$ for the gradients). (Left): MAE of the predicted energies. (Middle): MAE of the predicted QM gradients. (Right): MAE of the predicted MM gradients. For comparison, the purple lines denote the performance of DFTB on the validation set (dashed line) and test set (solid line).
    }\label{fig:performance_delta}
\end{figure}

As ML models use a locality ansatz, it is difficult for them to describe long-range interactions. Semi-empirical methods, on the other hand, are known to be able to account for long-range interactions. Thus, a $\Delta$-learning\cite{Ramakrishnan2015} scheme based on a semi-empirical method might be a valuable approach. In such a scheme, the ML model introduces a correction to the semi-empirical method to recover the accuracy of the higher-level method. Although this decreases naturally the computational efficiency of the production run, since an additional computation is required at each time step $t$, we will demonstrate in the following that the performance is increased significantly, and that the number of weight parameters can be reduced. As a result, the amount of training data points needed is reduced, the training procedure is accelerated, and the extrapolation capabilities of the model are possibly increased.
\newline
\newline
\begin{table}[H]
    \centering
    \scalebox{0.8}{
    \begin{tabular}{|c|c|cc|ccc|}
    \hline
    \textbf{Settings ID}  & $\Delta$-\textbf{Learning} & \multicolumn{2}{c|}{\textbf{Gradient correction}} & \multicolumn{3}{c|}{\textbf{Cutoff}}\\ 
    & & $\omega_0$ & $\omega_1$ & $R_{\text{sym,Rad}}^{\text{QM-QM}}$ [nm] &$R_{\text{sym,Rad}}^{\text{QM-MM}}$ [nm] & $R_c$ [nm] \\ \hline \hline
    M5    &Yes&0&0&1.0&1.0&0.6\\\hline
    M6    &Yes&1&1&1.0&1.0&0.6\\\hline
    M7    &Yes&1&200&1.0&1.0&0.6\\\hline
    \end{tabular}}
    \caption{Settings for the ML models used to test the influence of including gradient correction together with $\Delta$-learning. $\omega_0$ and $\omega_1$ are weight parameters for the gradient contribution (Eq. (\ref{eq:gradient_corrected})). $R_c$ is the cutoff radius for the MM partial charges in the QM/MM calculation to generate the data set, $R_{\text{sym,Rad}}^{\text{QM-QM}}$ is the cutoff for the QM-QM symmetry functions, and $R_{\text{sym,Rad}}^{\text{QM-MM}}$ the cutoff for the QM-MM symmetry functions. The test system is benzene in water.}
    \label{tab:ID2} 
\end{table}
Table \ref{tab:ID2} summarizes the settings. The $\Delta$-learning model M5 uses the same settings as M1, i.e. only two-body terms for the QM - MM interactions, no gradient correction ($\omega_0$=0, $\omega_1$=0), and a cutoff for the symmetry functions of 1.0\,nm. M6 uses the same settings as M3, i.e. gradient information are included in the training procedure ($\omega_0$=1, $\omega_1$=1). M7 uses stronger gradient regularization ($\omega_0=1$, $\omega_1=200$), and is thus comparable to M4. In Figure \ref{fig:performance_delta}, the $\Delta$-learning models are compared to M3.  

The results show that using gradient correction is crucial for performance as M3 (gradient correction but no $\Delta$-learning) clearly outperforms M5 ($\Delta$-learning but no gradient correction). At the same time, the prediction accuracy can be substantially increased by using a $\Delta$-learning in combination with gradient correction (M6). However, while the MAE with M6 is below 4.18~kJ~mol$^{-1}$ for the validation set, it is above it for the test set. By increasing the regularization (M7, $\omega_1=200$), the performance can be further improved for the test set. M7 presents thus the overall best performing model. In general, it is important to carefully monitor the performance of neural networks as a function of their parameters.

Using $\Delta$-learning and gradient correction improved the accuracy of all three properties, i.e. energies, QM gradients, and MM gradients, significantly already at a small number of neurons per hidden layer. This indicates that only a reduced number of training data points is necessary to achieve the accuracy of the reference method (see discussion below). It also suggests that the extrapolation capabilities of M6/M7 might be better than those of the previous models. This makes the model particularly interesting for use cases, where the ensemble of conformations/configurations of the system in the training set is not complete. 
To the best of our knowledge, the lowest level of theory available for such a $\Delta$-learning scheme is DFTB. Other semi-empirical methods do not incorporate the partial charges in an SCF manner in the Hamiltonian. Rather, they describe the interaction based on parameters. This means that the gradients on the MM particles usually do not correlate with those computed by the reference method (DFT).

\subsubsection{Size of the Training Set}
A major issue for learning of complex systems is that not all relevant configurations can be enumerated. Thus, ML approaches that can be trained on smaller data sets are especially valuable. As shown above, $\Delta$-learning requires less weight parameters to achieve chemical accuracy on the validation and test set of benzene in water, which indicates that less training points are needed compared to the ML models trained directly on the full energies and gradients. 
To quantify this observation, the M4 ($\omega_1=200$, 80 neurons per layer) and M7 ($\omega_1=200$, 10 neurons per layer) models were trained on only 10\% of the data set. This was done by changing the training/validation/test splitting from 70/20/10 to 10/20/10 (i.e. discarding the last 60\% of the training set). More precisely, the ML models with a reduced data set set are fitted on the first 1000 data points of the training set, while the validation and test set remain the same.
The results are presented in Table \ref{tab:10percent}.
It is striking that for M4 the prediction accuracy on the energies and gradients of the same validation and test sets drops by up to 50\%, when trained only on 10\% of the data set. For M7, on the other hand, the error on the energies remains relatively constant when only trained on 10\% of the data set. The performance on the gradients worsens as well but to a smaller extent.

\begin{table}[H]
    \centering
    \scalebox{0.75}{
    \begin{tabular}{|c|cc|cc|cc|}
    \hline
    \textbf{Model} & \multicolumn{2}{c|}{\textbf{Energy}} & \multicolumn{2}{c|}{\textbf{QM gradients}} & \multicolumn{2}{c|}{\textbf{MM gradients}} \\
         & \textbf{MAE} & \textbf{RMSE} & \textbf{MAE} & \textbf{RMSE} & \textbf{MAE} & \textbf{RMSE}   \\
         & [kJ\,mol$^{-1}$] & [kJ\,mol$^{-1}$] & [kJ\,mol$^{-1}$\,nm$^{-1}$] & [kJ\,mol$^{-1}$\,nm$^{-1}$] & [kJ\,mol$^{-1}$\,nm$^{-1}$] & [kJ\,mol$^{-1}$\,nm$^{-1}$] \\ \hline \hline
   M4 (70\%) &2.5, 1.8, 4.0 & 3.0, 2.2, 6.1 & 31.9, 34.0, 48.5 & 41.8, 45.9, 70.7 & 4.2, 5.2, 6.5 & 6.8, 8.5, 11.2 \\
   M4 (10\%) & 6.2, 3.9, 5.0 & 6.6, 4.8, 6.7 & 64.6, 89.7, 96.2 & 86.6, 122.0, 135.0 & 5.4, 7.1, 8.4  & 8.3, 12.3, 14.3 \\\hline
   M7 (70\%) & 1.8, 1.6, 3.2 & 2.2, 2.1, 4.6 & 23.7, 22.8, 29.4 & 31.1, 30.0, 44.2 &  3.5, 3.0, 4.7 & 6.4, 5.4, 8.9 \\
   M7 (10\%) &1.1, 1.6, 3.2 & 1.4, 2.0, 4.1 & 25.4, 31.1, 39.3 & 33.1, 41.7, 59.9 & 3.5, 3.5, 5.1 & 6.1, 6.2, 9.5\\\hline \hline
   DFTB & 6.0, 5.0 & 7.5, 6.6 & 143.2, 154.2 & 190.2, 209.3 & 6.0, 7.6 & 10.5, 14.1 \\
  \hline
    \end{tabular}}
    \caption{Comparison of the performance of M4 and M7 on the data set of benzene in water when trained on 10\% or 70\% of the data set. Performance is reported as the mean absolute error (MAE) and root-mean-square-error (RMSE). The first number indicates the performance on the training set, the second number that on the validation set, and the third number that on the test set. The reference data points were computed with B2-PLYP/def2-TZVP. For comparison, the performance of DFTB on the validation and test set is given. After the standard training protocol described in the Method section, M4 (10\%) and M7 (10\%) were trained for 50 more steps with a learning rate of $3.5\cdot 10^{-4}$.}
    \label{tab:10percent}
\end{table}

\subsubsection{Cutoff Size}
Fitting a ML model for the benzene in water system did not require the inclusion of long-range interactions as the partial charges beyond 0.6\,nm were truncated. To assess the effect of the cutoff size and thus the improved accounting for long-range interactions, the test system of uracil in water is used in the following. The cutoff $R_{c}$ is increased up to 1.4\,nm, which corresponds to 1500-2000 MM partial charges in the cutoff sphere. A cutoff of 1.4\,nm is used in the GROMOS force field,\cite{Schmid2011} of which the SPC/E water model is part of. Table \ref{tab:results_1.4nm} lists the settings M8-M13. All models use gradient correction, M11-M13 follow the $\Delta$-learning scheme with DFTB. 
M11 serves as a ``worst-case'' model for $\Delta$-learning, where the MM particles are simulated solely with the semi-empirical method, and the ML model introduces a local correction to the QM subsystem. 
\begin{table}[H]
    \centering
    \begin{tabular}{|c|ccc|c|}
    \hline
  \textbf{Settings ID}  & \multicolumn{3}{c|}{\textbf{Cutoff}} & \textbf{$\Delta$-Learning} \\
  & $R_{\text{sym,Rad}}^{\text{QM-QM}}$ [nm] & $R_{\text{sym,Rad}}^{\text{QM-MM}}$ [nm] & $R_{c}$ [nm] & \\ \hline \hline
  M8     & 0.52 & 0.00 & 1.40&No \\\hline
  M9     & 0.52 & 0.52 & 1.40&No\\\hline
  M10     & 1.40 & 1.40 & 1.40&No \\\hline
  M11    & 0.52 & 0.00 & 1.40&Yes\\\hline
  M12     & 0.52 & 0.52 & 1.40&Yes \\\hline
  M13     & 1.40 & 1.40 & 1.40&Yes \\\hline
    \end{tabular}
    \caption{Settings for the ML models used to test the influence of the cutoff size. All models employ only two-body terms for the QM - MM interactions and gradient correction. The models were trained with 10 neurons per hidden layer. $R_c$ is the cutoff radius for the MM partial charges in the QM/MM calculation, $R_{\text{sym,Rad}}^{\text{QM-QM}}$ is the cutoff for the QM-QM symmetry functions, and $R_{\text{sym,Rad}}^{\text{QM-MM}}$ the cutoff for the QM-MM symmetry functions. The test system is uracil in water.}
    \label{tab:results_1.4nm}
\end{table}

\begin{figure}[H]
    \centering
    \includegraphics[width=\textwidth]{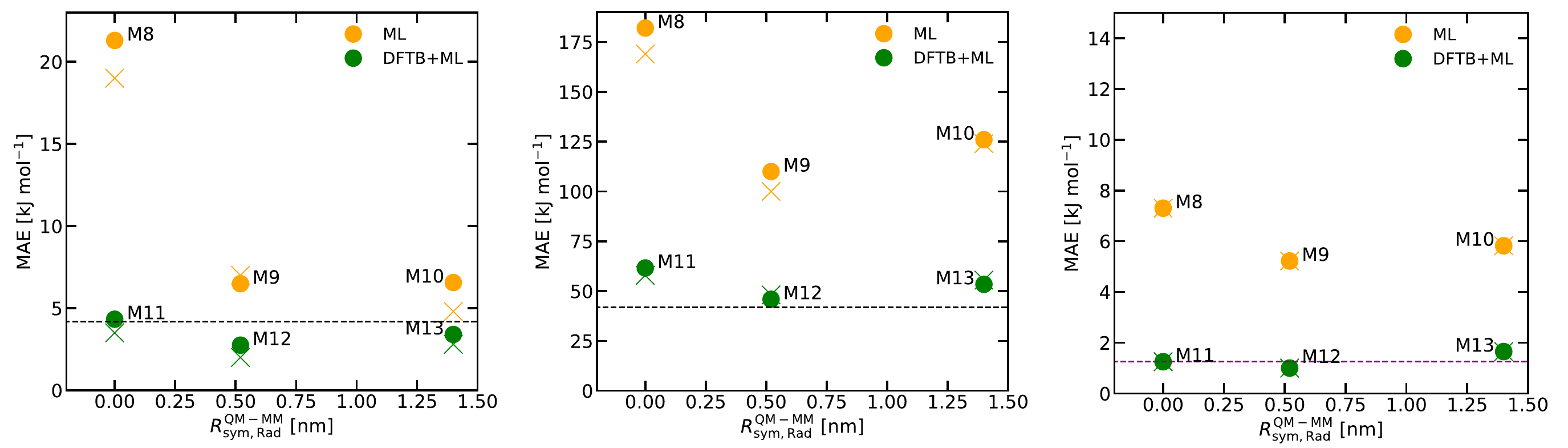}
    \caption{Performance of the settings M8-M10 (no $\Delta$-learning, yellow) and M11-M13 ($\Delta$-learning, green) on the validation (filled circles) and test (crosses) set of uracil in water as a function of the cutoff $R_{\text{sym,Rad}}^{\text{QM-MM}}$. The reference data points were computed with $\omega$B97X-D3/def2-TZVP. The performance is reported as the mean absolute error (MAE). The black dashed line indicates chemical accuracy (i.e. 4.18 kJ mol$^{-1}$ for the energies, and 41.8 kJ mol$^{-1}$ nm$^{-1}$ for the gradients). (Left): MAE of the predicted energies. (Middle): MAE of the predicted QM gradients. (Right): MAE of the predicted MM gradients. For comparison, the purple lines denote the performance of DFTB on the validation set (dashed line) and test set (solid line).}
    \label{fig:performance_delta2}
\end{figure}

The results with the models M8-M13 and the DFTB baseline for uracil in water are shown in Figure \ref{fig:performance_delta2} and summarized in Table \ref{tab:results_uracil}. 
We will first discuss the models without $\Delta$-learning, M8-M10. For these, long-range interactions need to be learned completely by the ML model. M8 has no information about the surrounding environment, and thus the performance on the energies and the QM gradients is poor. 
Increasing the cutoff of the symmetry function for the description of the MM environment to 0.52\,nm (M9) increases the performance significantly. The MAE on the energies on the validation set is reduced by nearly a factor 4, and the description of the gradients also becomes better. Interestingly, increasing the cutoff further worsens the results. The reason is likely that the training data set is relatively sparse (7000 data points). Increasing the cutoff in the symmetry functions also increases the number of weight parameters, which need to be fitted. This complicates the training procedure, making it harder to find meaningful minima. 
The situation is different for the $\Delta$-learning approach (M11-M13). Here, the semi-empirical method can be used to incorporate important long-range interactions, while the ML models serves as a local correction. A striking example provides M11. M11 like M8 has no information about the surrounding partial charges. Nevertheless, the model almost reaches chemical accuracy on the validation set. Compared to M8, the gradients on the QM particles are improved by a factor three, and the gradients on the partial charges by a factor 7. Including information about the partial charges in the surrounding up to 0.52\,nm improves the results further. Again, going from a cutoff of 0.52\,nm to 1.4\,nm worsens the results, likely due to the same reason. However, given a sufficiently large training set (larger than in the current cases), a $R_{\text{sym,Rad}} > 0.52$~nm may be desirable. 

Lastly, we compare DFTB to the ML models. All ML models outperform DFTB on predicting the QM gradients. However, DFTB outperforms the ML models without $\Delta$-learning on the MM gradients. This highlights again the importance of applying $\Delta$-learning for systems, where long-range interactions are important and only sparse data sets are available. Further, it is interesting to observe that DFTB outperforms M8, which has no information about the surrounding partial charges, on the QM energies. This finding suggests that well-fitted semi-empirical methods may be preferred over ML models, if the latter do not account appropriately for the surrounding partial charges.

\begin{table}[H]
    \centering
    \scalebox{0.75}{
    \begin{tabular}{|c|cc|cc|cc|}
    \hline
    \textbf{Model} & \multicolumn{2}{c|}{\textbf{Energy}} & \multicolumn{2}{c|}{\textbf{QM gradients}} & \multicolumn{2}{c|}{\textbf{MM gradients}} \\
         & \textbf{MAE} & \textbf{RMSE} & \textbf{MAE} & \textbf{RMSE} & \textbf{MAE} & \textbf{RMSE}   \\
         & [kJ\,mol$^{-1}$] & [kJ\,mol$^{-1}$] & [kJ\,mol$^{-1}$\,nm$^{-1}$] & [kJ\,mol$^{-1}$\,nm$^{-1}$] & [kJ\,mol$^{-1}$\,nm$^{-1}$] & [kJ\,mol$^{-1}$\,nm$^{-1}$] \\ \hline \hline
  M8  & 16.2, 21.3, 19.0 & 20.0, 26.3, 23.5 & 128.0, 182.0, 169.0 & 170.0, 278.0, 242.0 & 7.8, 7.3, 8.1 & 23.3, 23.5, 24.6\\\hline
  M9  & 5.2, 6.5, 7.0 & 6.2, 8.5, 8.6 & 69.9, 110.0, 100.0 & 91.6, 176.0, 139.0 & 5.6, 5.2, 5.89 & 10.7, 10.1, 11.1 \\\hline
  M10 & 2.9, 6.6, 4.8 &  3.7, 8.5, 6.0 & 63.8, 126.0, 124.0 & 81.7, 191.0, 181.0 & 6.2, 5.8, 6.5 &11.6, 10.9, 12.0 \\\hline \hline
  
M11    &4.6, 4.3, 3.5 & 5.8, 5.5, 4.2 &47.1, 61.6, 57.9 & 60.6, 81.8, 76.5 & 1.3, 1.3, 1.3 & 5.2, 5.5, 5.0\\\hline
M12    & 1.2, 2.8, 2.0 & 1.5, 3.6, 2.7 &34.3, 45.9, 48.1 &44.6, 62.0, 64.4 & 1.1, 1.0, 1.0 & 3.2, 3.2, 3.1 \\\hline
M13    & 0.9, 3.4, 2.8 &1.2, 4.2, 3.5 &29.8, 53.4, 55.5 & 38.4, 72.2, 75.4 &1.7, 1.7, 1.7 & 3.7, 3.9, 3.8 \\\hline \hline
   DFTB & 12.4, 12.7 & 16.0, 15.2 & 160.0, 151.5 & 700.1, 716.9 & 1.3, 1.3 & 5.5, 5.0  \\\hline
    \end{tabular}}
    \caption{Performance of the models M8-M13 for uracil in water. The performance is reported as the mean absolute error (MAE) and root-mean-square-error (RMSE). The first value indicates the performance on the training set (70\%), the second number on the validation set (20\%), and the third number on the test set (10\%). The reference data points were computed with $\omega$B97X-D3/def2-TZVP. For comparison, the performance of DFTB on the validation and test set is given.}
    \label{tab:results_uracil}
\end{table}

To assess the transferability of the conclusions from the previous test systems, we analyzed the performance of the settings M11-M13 on a third test system, the (close to) transition state of the S$_N$2 reaction between CH$_3$Cl and Cl$^{-}$. For this system, larger changes in the electronic structure within the configurational ensemble are expected. Furthermore, as it is a charged system, it is interesting to see, how well the long-range interactions are described by the $\Delta$-learning models. %
The results for M11-M13 for the CH$_3$Cl/Cl$^{-}$ test system are summarized in Table \ref{tab:performance_chlormethane}. In general, we see the same behaviour as for uracil in water, i.e. a cutoff of 0.52\,nm performs best. 

\begin{table}[H]
    \centering
    \scalebox{0.8}{
    \begin{tabular}{|c|cc|cc|cc|}
    \hline
    \textbf{Model} & \multicolumn{2}{c|}{\textbf{Energy}} & \multicolumn{2}{c|}{\textbf{QM gradients}} & \multicolumn{2}{c|}{\textbf{MM gradients}} \\
         & \textbf{MAE} & \textbf{RMSE} & \textbf{MAE} & \textbf{RMSE} & \textbf{MAE} & \textbf{RMSE}   \\
         & [kJ\,mol$^{-1}$] & [kJ\,mol$^{-1}$] & [kJ\,mol$^{-1}$\,nm$^{-1}$] & [kJ\,mol$^{-1}$\,nm$^{-1}$] & [kJ\,mol$^{-1}$\,nm$^{-1}$] & [kJ\,mol$^{-1}$\,nm$^{-1}$] \\ \hline \hline
   M11  & 4.6, 3.7, 3.5 & 6.3, 4.6, 5.0 & 59.9, 56.5, 45.5 & 94.1, 83.9, 86.4 & 1.2, 1.1, 1.1 & 6.5, 5.2, 5.2 \\\hline
   M12     & 1.8, 1.7, 2.2 & 2.3, 2.1, 2.8 &24.9, 29.5, 30.7& 33.6, 40.8, 44.2& 1.0, 1.0, 1.0 & 3.9, 3.3, 3.0 \\\hline
   M13 & 1.2, 2.3, 4.5& 1.5, 2.9, 4.9 & 25.8, 40.0, 43.7& 33.6, 54.4, 67.7& 1.5, 1.4, 1.4& 4.2, 3.8, 3.6 \\\hline \hline
    DFTB  &6.6, 6.8&8.4, 9.4& 191.6, 203.9& 236.6, 276.9& 1.1, 1.1  &5.2, 5.2\\\hline
  \end{tabular}}
    \caption{Performance of the models M11-M13 for the (close to) transition state of CH$_3$Cl/Cl$^{-}$ in water. The performance is reported as the mean absolute error (MAE) and root-mean-square-error (RMSE). The first value indicates the performance on the training set (70\%), the second number on the validation set (20\%), and the last number on the test set (10\%). The reference data points were computed with Head-Gordon's range separated functional $\omega$B97X-D3. For comparison, the performance of DFTB on the validation and test set is given.}
    \label{tab:performance_chlormethane}
\end{table}

\subsection{Application of ML Models in MD Simulations}
So far we have assessed the performance of the ML models in terms of MAE (and RMSE) on a validation and test set. However, these results can be to some degree misleading because rare outliers get averaged. Such rare outliers might be less tolerable in the context of an actual MD simulations, where the results of the next step depend directly on the results of the previous step.

\begin{figure}[H]
    \centering
    \includegraphics[width=\textwidth]{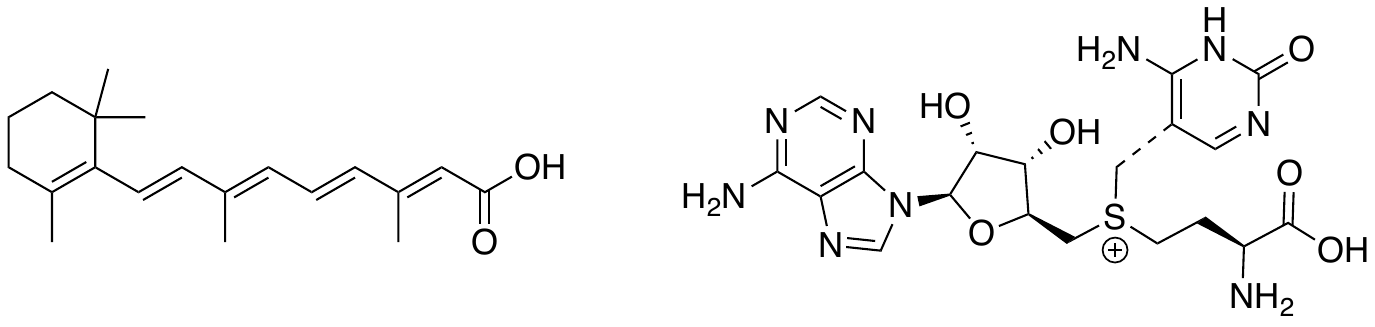}
    \caption{Two larger test systems used for MD simulations. (Left): Retinoic acid in water (50 QM atoms and about 2500 MM partial charges with a cutoff of 1.4\,nm). (Right): Transition state of the chemical reaction between SAM and cytosine in water (63 QM atoms and about 3500 MM partial charges with a cutoff of 1.4\,nm).}
    \label{fig:large_systems}
\end{figure}

In order to judge the usefulness of HDNNP models in practice, it is important to test their performance in actual MD simulations. For this purpose, we have chosen two test systems with larger conformational flexibility: (i) retinoic acid (a vitamin A derivative) in water, and (ii) the (close to) transition state of the reaction of SAM with cytosine in water (Figure \ref{fig:large_systems}). Sampling to generate the training and validation sets was performed at elevated temperature (400\,K) with a time step of 0.5\,fs for both systems. Instead of an external test set, MD simulations were carried out at $T=298$\,K. Our main objective in this section is to show that given a sparse training set the proposed $\Delta$-learning scheme is able to produce stable trajectories, which describe the PES sufficiently well even for unseen data points. We therefore use as above only 10'000 data points for training, even though the complexity of the systems is much higher compared to the previously studied ones. For real case studies, e.g. the computation of free-energy profiles, a more sophisticated sampling procedure for the training set is recommended.

The ML models investigated employ the settings M12, i.e. $\Delta$-learning scheme with DFTB as baseline, 10 neurons per hidden layer, only two-body terms for the QM - MM interactions, gradient correction (with $\omega_0=1$ and $\omega_1=200$), and $R_{\text{sym,Rad}}$ of 0.52\,nm.
The QM/MM reference calculations were computed with a cutoff $R_c$ of 1.40\,nm.

\subsubsection{Retinoic Acid in Water}
\begin{figure}[H]
   \centering
    \includegraphics[width=0.8\textwidth]{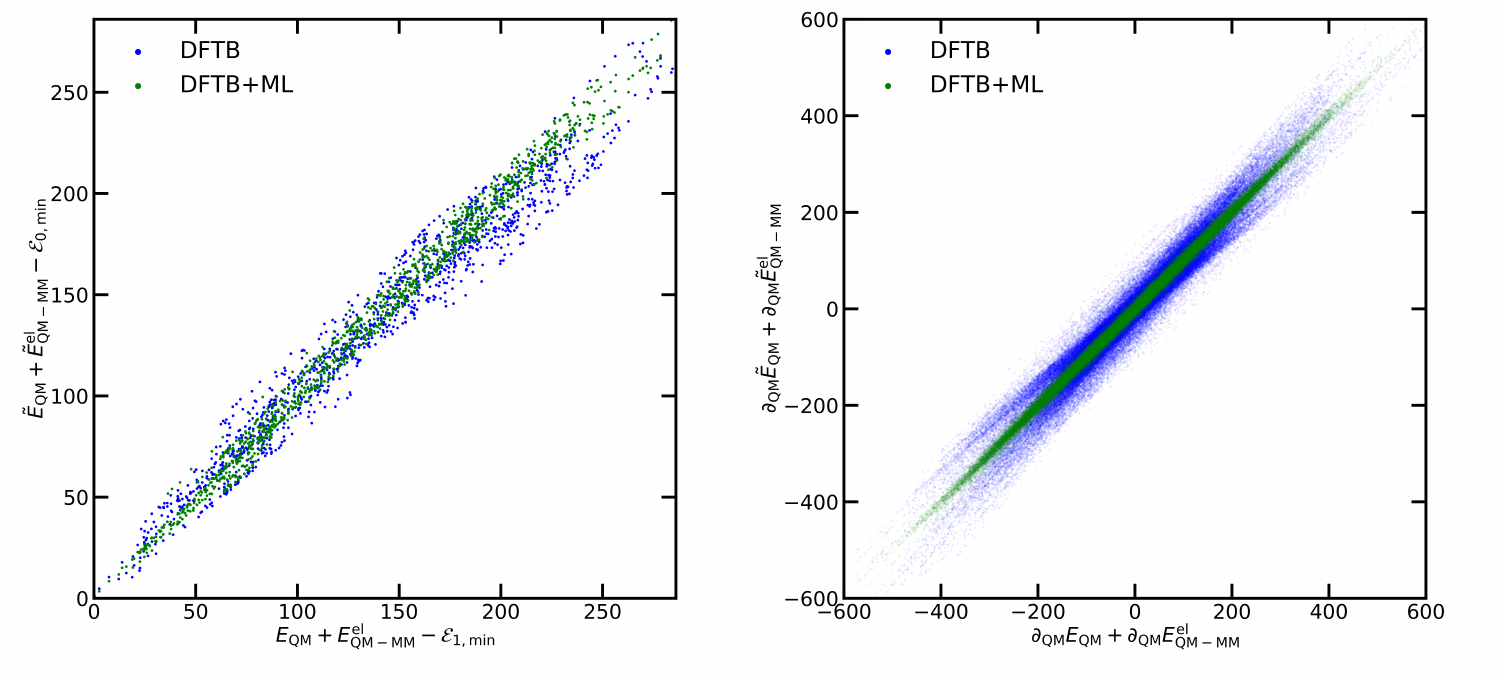}
    \caption{Comparison between the reference DFT method and the DFTB Hamiltonian (blue) and the DFTB + ML model (green) on the validation set of retinoic acid in water. The reference data points were computed with BP86/def2-TZVP. (Left): Energies, units in kJ mol$^{-1}$. (Right): QM gradients, units in kJ mol$^{-1}$ nm$^{-1}$. We define $\frac{\partial}{\partial \vec{R}^{\text{QM}}} \equiv \partial_\text{QM}$. The minimum energy for the reference and the ML prediction is subtracted, as indicated with $-\epsilon_{0,\text{min}}$ and $-\epsilon_{1,\text{min}}$.}
    \label{fig:retinacid_results} 
\end{figure}

First, we assessed the performance of the ML model on a validation set, as done in the previous sections. Table \ref{tab:retinoic_acid_results} summarizes the results in terms of MAE and RMSE, Figure \ref{fig:retinacid_results} shows them graphically. 
Using the $\Delta$-learning model, the performance is improved compared to the DFTB baseline for all three properties, including the MM gradients.
\begin{table}[H]
    \centering
    \scalebox{0.8}{
    \begin{tabular}{|c|cc|cc|cc|}
    \hline
    \textbf{Model} & \multicolumn{2}{c|}{\textbf{Energy}} & \multicolumn{2}{c|}{\textbf{QM gradients}} & \multicolumn{2}{c|}{\textbf{MM gradients}} \\
         & \textbf{MAE} & \textbf{RMSE} & \textbf{MAE} & \textbf{RMSE} & \textbf{MAE} & \textbf{RMSE}   \\
         & [kJ\,mol$^{-1}$] & [kJ\,mol$^{-1}$] & [kJ\,mol$^{-1}$\,nm$^{-1}$] & [kJ\,mol$^{-1}$\,nm$^{-1}$] & [kJ\,mol$^{-1}$\,nm$^{-1}$] & [kJ\,mol$^{-1}$\,nm$^{-1}$] \\ \hline \hline
   ML & 3.9, 4.4& 4.4, 5.0& 43.8, 44.8& 59.7, 61.8 &  1.1, 1.1& 3.3, 3.1
 \\\hline
    DFTB & 9.9 & 12.6 & 216.3 & 286.7 & 1.3 &4.5 \\\hline
  \end{tabular}}
    \caption{Performance of the $\Delta$-learning model on the training and validation set of retinoic acid in water. The performance is reported as the mean absolute error (MAE) and root-mean-square-error (RMSE). The first value corresponds to the training set, the second value to the performance on the validation set. The reference data points were computed with BP86/def2-TZVP. For comparison, the performance of DFTB on the validation set is given.}
    \label{tab:retinoic_acid_results}
\end{table}

\begin{figure}[H]
   \centering
    \includegraphics[width=0.8\textwidth]{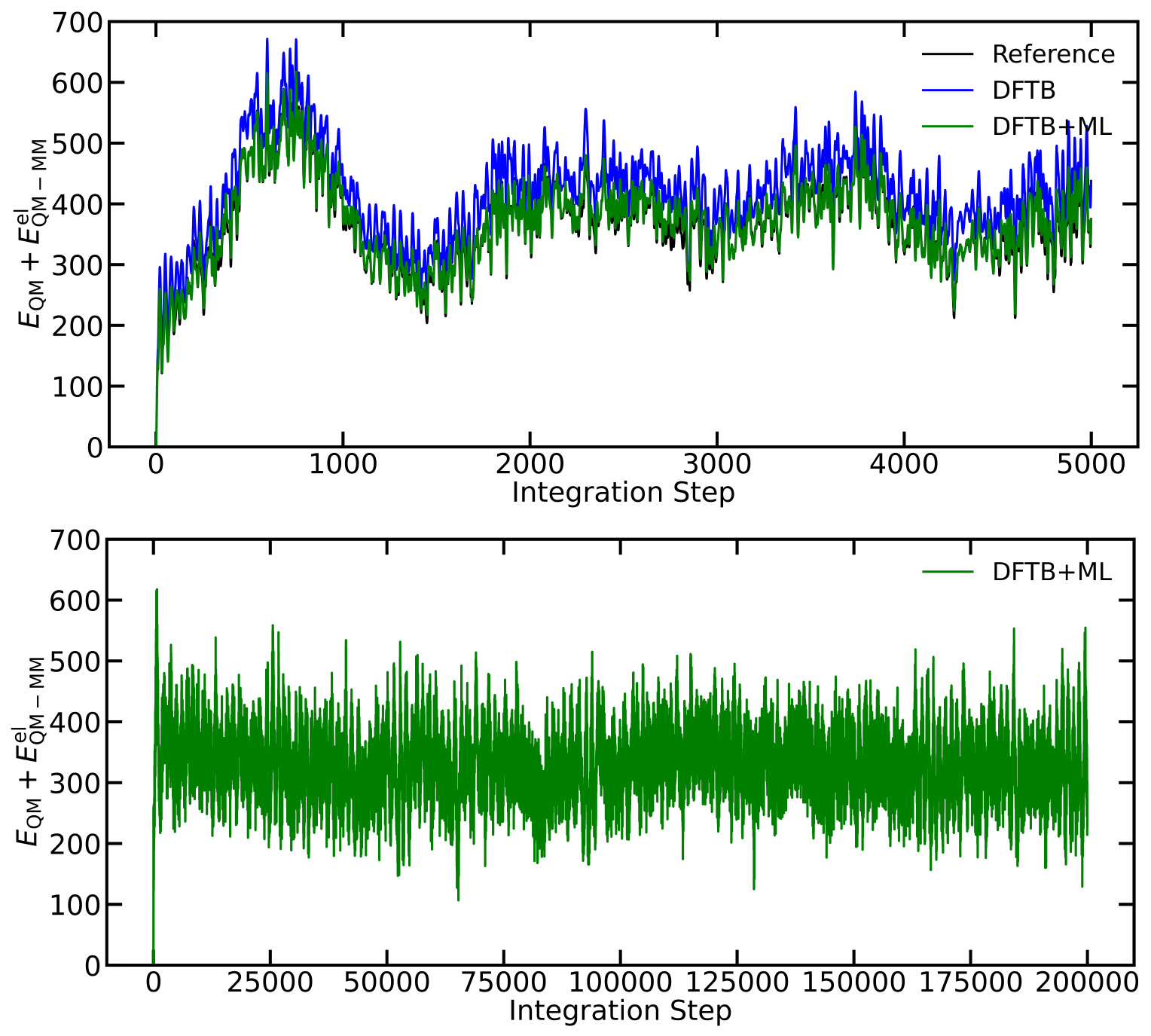}
    \caption{MD simulation of retinoic acid in water (integration size 0.5\,fs). (Top): Energy ($E_\text{QM}+E_\text{QM--MM}^{\text{el}}$) trajectory from 5000 consecutive steps performed by the DFTB + ML model (green). In each step, the energies were also computed with DFTB (blue) and the DFT reference (BP86/def2-TZVP, black) for comparison. The trajectories are referenced to their zero energy. The energy deviation from the reference over the complete trajectory with respect to the minimum energy is: MAE = 5.8 kJ mol$^{-1}$ and RMSE = 7.6 kJ mol$^{-1}$ for the  DFTB + ML model, and MAE = 52.8 kJ mol$^{-1}$ and RMSE = 54.3 kJ mol$^{-1}$ for DFTB. 
    (Bottom): Energy trajectory from 200'000 consecutive steps performed by the DFTB + ML model (green).} 
    \label{fig:MD_1} 
\end{figure}

Next, we tested the trained ML model by performing an (QM)ML/MM MD simulation for 5000 steps (top panel in Figure \ref{fig:MD_1}). In order to compare the ML-corrected energies with those of the DFT reference and the DFTB baseline simulation, single point calculations were performed for each configuration in the ML + DFTB trajectory. As shown in Figure \ref{fig:MD_1}, the energies of ML + DFTB agree better with the DFT reference than the DFTB baseline. Although the latter performs relatively well, one should keep in mind that the deviations from the reference method accumulate along an MD trajectory. Thus, even relatively small deviations might result in largely different trajectories when starting from the same coordinates. We note that the MAE on the trajectory is with 5.8\,kJ\,mol$^{-1}$ larger than the MAE of 3.9\,kJ\,mol$^{-1}$ and 4.4\,kJ\,mol$^{-1}$ for the training and validation sets, respectively. 
The reason for this is that the starting point of the trajectory is the relaxed structure, a configuration never seen by the HDNNP. To achieve a higher accuracy in practical MD simulations for complex molecules such as retinoic acid, more training data points will be needed for the HDNNP. To assess the stability of the ML model, we extended the simulation to 200'000 steps (bottom panel in Figure \ref{fig:MD_1}).

\subsubsection{SAM/Cytosine Transition State in Water}
\begin{figure}
    \centering
    \includegraphics[width=0.8\textwidth]{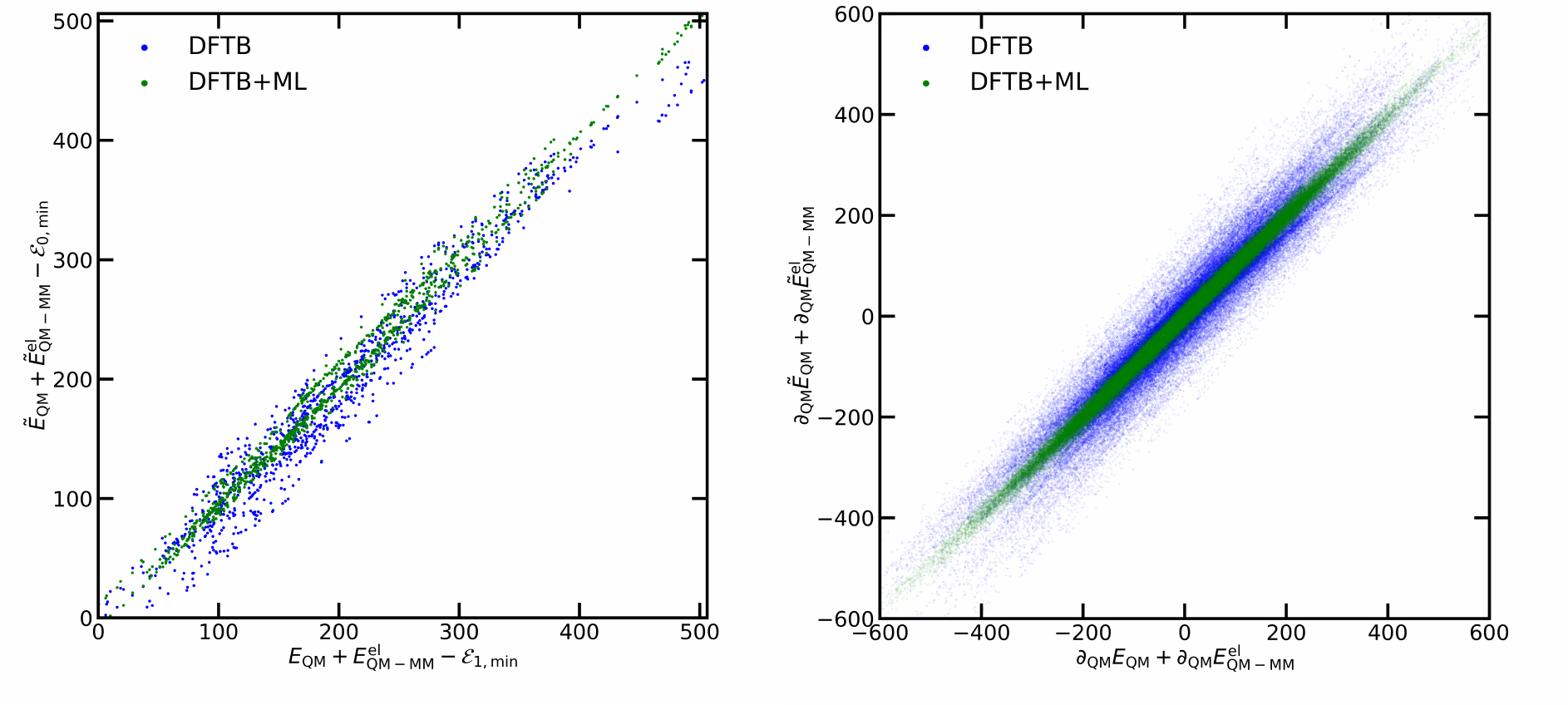}
    \caption{Comparison between the reference DFT method and the DFTB Hamiltonian (blue) and the DFTB + ML  model (green) on the validation set of the (close to) transition state of SAM/cytosine in water. (Left): Energies, units in kJ mol$^{-1}$. (Right): QM gradients, units in kJ mol$^{-1}$ nm$^{-1}$. We define $\frac{\partial}{\partial \vec{R}^{\text{QM}}} \equiv \partial_\text{QM}$. The minimum energy for the reference and the ML prediction is subtracted, as indicated with $-\epsilon_{0,\text{min}}$ and $-\epsilon_{1,\text{min}}$.}
    \label{fig:SAM_correlation} 
\end{figure}

In biochemistry, S-adenosylmethionate (SAM) is a co-factor for the transfer of a methyl group by enzymes (methyltransferases). Here, we investigated the transition state of the chemical reaction between SAM and cytosine. Again, we first assessed the performance on a validation set as done above (Table \ref{tab:results_SAM} and Figure \ref{fig:SAM_correlation}). The $\Delta$-learning model clearly outperforms the DFTB baseline. 

\begin{table}[H]
    \centering
    \scalebox{0.8}{
    \begin{tabular}{|c|cc|cc|cc|}
    \hline
    \textbf{Model} & \multicolumn{2}{c|}{\textbf{Energy}} & \multicolumn{2}{c|}{\textbf{QM gradients}} & \multicolumn{2}{c|}{\textbf{MM gradients}} \\
         & \textbf{MAE} & \textbf{RMSE} & \textbf{MAE} & \textbf{RMSE} & \textbf{MAE} & \textbf{RMSE}   \\
         & [kJ\,mol$^{-1}$] & [kJ\,mol$^{-1}$] & [kJ\,mol$^{-1}$\,nm$^{-1}$] & [kJ\,mol$^{-1}$\,nm$^{-1}$] & [kJ\,mol$^{-1}$\,nm$^{-1}$] & [kJ\,mol$^{-1}$\,nm$^{-1}$] \\ \hline \hline
   ML & 6.3, 8.5& 6.9, 9.7 & 74.8, 74.6 &101.0, 100.0 & 2.3, 2.3 & 6.7, 6.7  \\\hline
   DFTB &17.1 & 21.9 & 263.7 & 377.2&2.3&7.6 \\\hline
  \end{tabular}}
    \caption{Performance of the $\Delta$-learning model on the training and validation sets of the (close to) transition state of SAM and cytosine in water. The performance is reported as the mean absolute error (MAE) and root-mean-square-error (RMSE). The first value corresponds to the training set, the second value to the performance on the validation set. The reference data points were computed with BP86/def2-TZVP. For comparison, the performance of DFTB on the validation set is given.} 
    \label{tab:results_SAM}
\end{table}

\begin{figure}[H]
   \centering
    \includegraphics[width=0.8\textwidth]{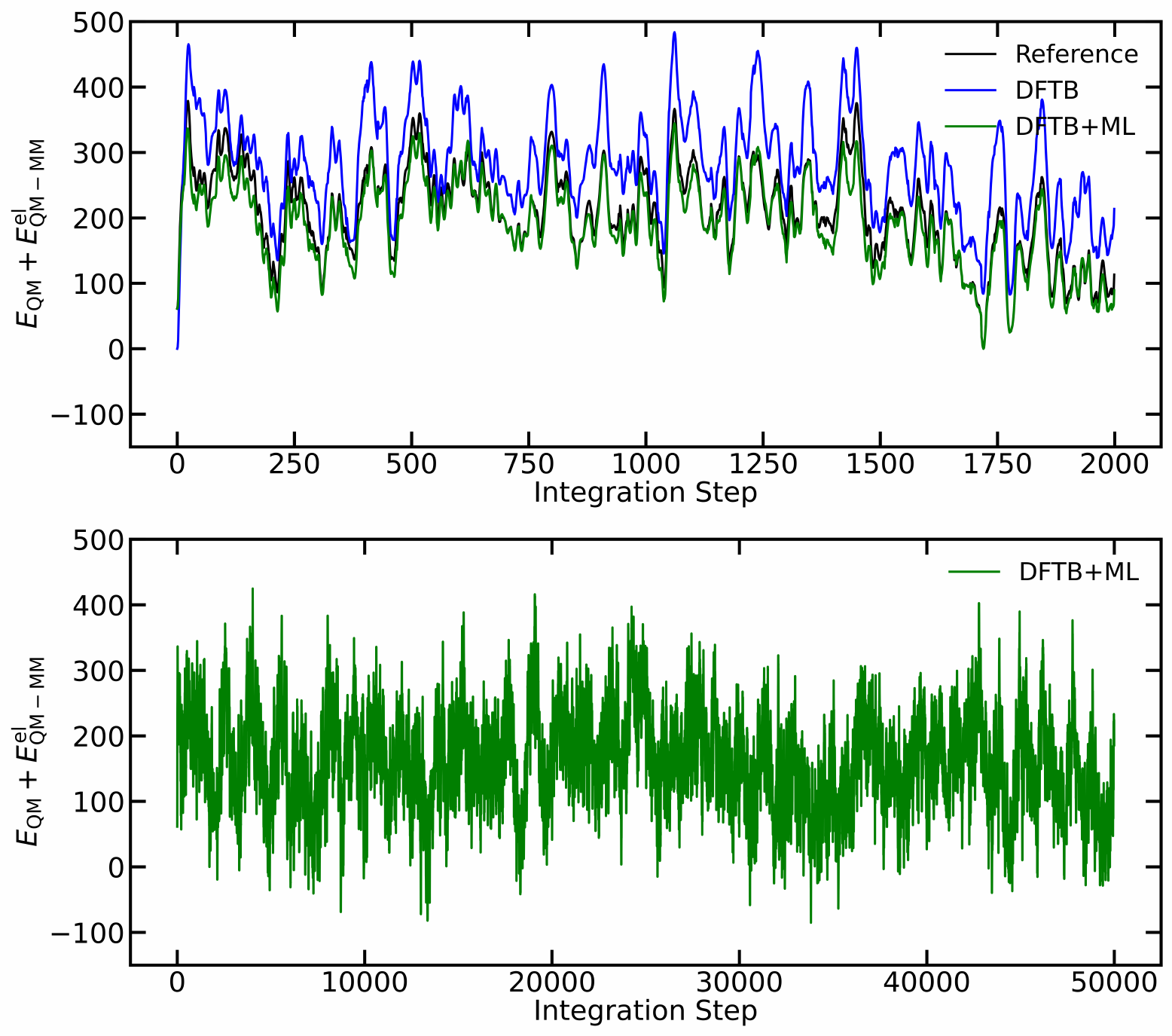}
    \caption{MD simulation of the SAM/cytosine transition state in water (integration size 0.5\,fs). (Top): Energy ($E_\text{QM}+E_\text{QM--MM}^{\text{el}}$) trajectory from 2'000 consecutive steps performed by the DFTB + ML model (green). In each step, the energies were also computed with DFTB (blue) and the DFT reference (BP86/def2-TZVP, black) for comparison. The energy deviation from the reference over the complete trajectory with respect to the minimum energy is: MAE = 18.1\,kJ\,mol$^{-1}$ and RMSE = 22.2\,kJ\,mol$^{-1}$ for the DFTB + ML model, and MAE = 74.9\,kJ\,mol$^{-1}$ and RMSE = 79.4\,kJ\,mol$^{-1}$ for DFTB. (Bottom): Energy trajectory from 50'000 consecutive steps performed by the DFTB + ML model (green).} 
    \label{fig:MD_2} 
\end{figure}

Next, we performed a (QM)ML/MM MD simulation for 2000 steps using the fitted model (top panel in Figure \ref{fig:MD_2}). As for the test system with retinoic acid in water, the energies with DFTB + ML agreed well with the DFT reference, and outperformed the DFTB baseline as indicated by the MAE and RMSE over the trajectory. To assess the stability of the model, we extended the simulation to 50'000 steps (bottom panel in Figure \ref{fig:MD_2}).
For both test systems, it was possible to carry out a stable (QM)ML/MM MD simulation for the selected number of steps (up to 200'000 steps). Of course, the possibility cannot be excluded at this point that a configuration, which is not well represented by the ML model, might be encountered in even longer simulations. 

\subsection{General Discussion}
The $\Delta$-learning scheme requires an explicit computation with the lower-level method at each time step. In this work, we have used DFTB as the lower-level method. For the SAM/cytosine system with 63 QM atoms and up to 3500 MM particles in the cutoff sphere, one evaluation with DFTB required less than a second (single core), while the corresponding DFT computation takes 60 to 80 minutes (on 4 cores, CPU time), which is more than three orders of magnitude longer. Thus, the additional time needed for the DFTB calculation at each step of the $\Delta$-learning scheme is less expensive than e.g. the pairlist creation in the MD engine.
An obvious limitation of the $\Delta$-learning approach are configurations, for which the DFTB PES is substantially different from the reference DFT PES. In such cases, the DFTB computation will converge very slowly (or not at all), limiting the usage. However, we did not encounter such an issue in the presented MD simulations.

\section{Summary and Conclusion}
In this work, we investigated the use of HDNNP in (QM)ML/MM MD simulations of condensed-phase systems with DFT (or $ab$ $initio$) accuracy for the QM subsystem. Standard QM/MM MD simulations at this level of accuracy are very expensive and only applicable to small systems. Using semi-empirical methods to describe the QM parts is much faster but also less accurate. In the HDNNP, the MM partial charges up to a certain cutoff are integrated as additional element type. 

We assessed the influence of different parameters on the prediction accuracy of energies, gradients of the QM particles, and gradients of the MM particles. First, we used a simple test system of an apolar, rigid solute in water. Three-body terms to describe the QM - MM interactions could be neglected, however, the inclusion of a gradient correction in the training procedure of the ML model was crucial to obtain sufficient accuracy on the gradients and energies. In general, we found that strong regularization was necessary during the training procedure to predict these gradients with decent accuracy. This may no longer apply when larger training sets are used.
As an alternative, we employed a $\Delta$-learning scheme, where the lower-level method is DFTB. Semi-empirical methods are known to handle long-range interactions well. 
For uracil in water, we showed that $\Delta$-learning allows one to decrease the cutoff of the symmetry functions substantially, while still outperforming classical HDNNP. The removal of symmetry functions reduces the number of weight parameters and the complexity of the training procedure. Smaller models usually require less training data points and display improved extrapolation capabilities. This may be a key factor for large biomolecular systems, where it is not feasible to enumerate all conformations/configurations for the training set.

The final $\Delta$-learning model was tested further by performing actual (QM)ML/MM MD simulations of relatively large systems, i.e. retinoic acid in water and the transition state of the chemical reaction between SAM and cytosine. Stable trajectories were obtained with an accuracy close to the DFT reference. The performance can be further boosted by a more sophisticated sampling strategy for the training and validation set. We envision that the results and findings presented in this study will enable the use of (QM)ML/MM MD simulations in practical applications.

\section*{Acknowledgements}
The authors thank Patrick Bleiziffer and Annick Renevey for helpful discussions. 
S.R. gratefully acknowledges financial support by the Swiss National Science Foundation (Grant Number 200021-178762) and by ETH Zurich (ETH-34 17-2). 

\bibliographystyle{b2}
\bibliography{references}

\end{document}